\input{psfig.sty}

\documentstyle{mn}

\newif\ifAMStwofonts
\AMStwofontstrue

\ifoldfss
  \ifCUPmtlplainloaded \else
    \NewTextAlphabet{textbfit} {cmbxti10} {}
    \NewTextAlphabet{textbfss} {cmssbx10} {}
    \NewMathAlphabet{mathbfit} {cmbxti10} {} % for math mode
    \NewMathAlphabet{mathbfss} {cmssbx10} {} %  "   "    "
  \fi
  \ifAMStwofonts
    \ifCUPmtlplainloaded \else
      \NewSymbolFont{upmath} {eurm10}
      \NewSymbolFont{AMSa} {msam10}
      \NewMathSymbol{\upi}     {0}{upmath}{19}
      \NewMathSymbol{\umu}     {0}{upmath}{16}
      \NewMathSymbol{\upartial}{0}{upmath}{40}
      \NewMathSymbol{\leqslant}{3}{AMSa}{36}
      \NewMathSymbol{\geqslant}{3}{AMSa}{3E}

    \fi
  \fi
\fi % End of OFSS

\ifnfssone
  \newmathalphabet{\mathit}
  \addtoversion{normal}{\mathit}{cmr}{m}{it}
  \addtoversion{bold}{\mathit}{cmr}{bx}{it}
  \newmathalphabet{\mathbfit} % math mode version of \textbfit{..}
  \addtoversion{normal}{\mathbfit}{cmr}{bx}{it}
  \addtoversion{bold}{\mathbfit}{cmr}{bx}{it}
  \newmathalphabet{\mathbfss} % math mode version of \textbfss{..}
  \addtoversion{normal}{\mathbfss}{cmss}{bx}{n}
  \addtoversion{bold}{\mathbfss}{cmss}{bx}{n}
  \ifAMStwofonts
    \ifCUPmtlplainloaded \else
      \UseAMStwoboldmath
      \makeatletter
      \new@mathgroup\upmath@group
      \define@mathgroup\mv@normal\upmath@group{eur}{m}{n}
      \define@mathgroup\mv@bold\upmath@group{eur}{b}{n}
      \edef\UPM{\hexnumber\upmath@group}
      \new@mathgroup\amsa@group
      \define@mathgroup\mv@normal\amsa@group{msa}{m}{n}
      \define@mathgroup\mv@bold\amsa@group{msa}{m}{n}
      \edef\AMSa{\hexnumber\amsa@group}
      \makeatother
      \mathchardef\upi="0\UPM19
      \mathchardef\umu="0\UPM16
      \mathchardef\upartial="0\UPM40
      \mathchardef\leqslant="3\AMSa36
      \mathchardef\geqslant="3\AMSa3E
    \fi
  \fi
\fi % End of NFSS release 1

\ifnfsstwo
  \DeclareMathAlphabet{\mathbfit}{OT1}{cmr}{bx}{it}
  \SetMathAlphabet\mathbfit{bold}{OT1}{cmr}{bx}{it}
  \DeclareMathAlphabet{\mathbfss}{OT1}{cmss}{bx}{n}
  \SetMathAlphabet\mathbfss{bold}{OT1}{cmss}{bx}{n}
  \ifAMStwofonts
    \ifCUPmtlplainloaded \else
      \DeclareSymbolFont{UPM}{U}{eur}{m}{n}
      \SetSymbolFont{UPM}{bold}{U}{eur}{b}{n}
      \DeclareSymbolFont{AMSa}{U}{msa}{m}{n}
      \DeclareMathSymbol{\upi}{0}{UPM}{"19}
      \DeclareMathSymbol{\umu}{0}{UPM}{"16}
      \DeclareMathSymbol{\upartial}{0}{UPM}{"40}
      \DeclareMathSymbol{\leqslant}{3}{AMSa}{"36}
      \DeclareMathSymbol{\geqslant}{3}{AMSa}{"3E}
    \fi
  \fi
\fi % End of NFSS release 2

\ifCUPmtlplainloaded \else
  \ifAMStwofonts \else % If no AMS fonts
    \def\upi{\pi}
    \def\umu{\mu}
    \def\upartial{\partial}
  \fi
\fi

\title[Dynamics of rotating stellar systems]{Dynamical evolution of 
        rotating stellar systems: \\
        I. Pre-collapse, equal mass system}
\author[Ch. Einsel and R. Spurzem]
       {Christian Einsel$^{1,2}$ and Rainer Spurzem$^2$ \\
        $^1$Institut f{\"u}r Astronomie und Astrophysik der 
            Universit{\"a}t Kiel,
        Olshausenstr. 40, D-24098 Kiel, Fed. Rep. of Germany \\
        $^2$Astronomisches Rechen-Institut, M{\"o}nchhofstr. 12-14,
            D-69115 Heidelberg, Fed. Rep. of Germany}
\date{Accepted .
      Received ;
      in original form }

\pagerange{\pageref{firstpage}--\pageref{lastpage}}
\pubyear{1997}

\newcommand{\D}{\displaystyle}

\begin{document}

\maketitle

\label{firstpage}

\begin{abstract}
The influence of rotation on the dynamical evolution of collisional
stellar systems is investigated by solving the orbit-averaged
Fokker-Planck equation in $(E,J_{z})$-space. We find that
large amounts of initial rotation drive the system into a phase of strong
mass loss while it is moderately contracting. The core is rotating even
faster than before although angular momentum is transported outwards.
At the same time the core is heating. Given these features
this phase can be associated with the gravo-gyro 'catastrophe' found by
Hachisu (1979). The increase in central angular momentum levels off after 
about 2-3 initial half-mass relaxation times indicating that the source 
of this `catastrophe' is depleted. Finally,
the central angular velocity increases again, but with a rather small power
of the central density -- the same power as for the central velocity dispersion
during self-similar contraction towards the collapse.
The rotation curves flatten and the ellipticity variations decrease with time,
but their shapes are very similar. These results suggest the existence
of a self-similar solution for a rotating cluster as well. The maximum 
values of rotational velocity and ellipticity occur at about the half 
mass radius.

\end{abstract}


\section{Introduction}

The study of the dynamical 
behaviour of spherically symmetric, collisional stellar systems has made 
considerable progress during the last decade. Numerical improvements
and refinements are applied to current techniques (Fokker-Planck, 
Monte-Carlo, fluid-dynamical or N-Body) in order to investigate anisotropy
and the post-collapse phase (Bett\-wieser \& Sugimoto 1984, Murphy {\em et al.}\
1990, Heggie \& Ramamani 1989, Takahashi 1996a,b, Spurzem 1994, 1996), 
stochastic energy input due to binaries (Giersz 1996), small-N statistics 
(Giersz \& Heggie 1994ab, Heggie 1996), primordial binaries (Hut 1996),
tidal effects, stellar evolution or mass spectra (Chernoff \& Weinberg 1990,
Lee {\em et al.}\ 1991, Fukushige \& Heggie 1995, Drukier 1995, Giersz \& Heggie
1996, 1997).  Considering anisotropy, there are still other features in 
addition to its hitherto studied effects on the radial variations of density
and velocity dispersion, one important effect being the feasibility to
flatten those systems. Binney (1978) applied the tensor virial theorem to
elliptical galaxies constructed from similar ellipsoids and
comprising flat rotation curves, and he derived formulae relating ellipticity
to the ratio of rotation velocity and velocity dispersion. 
The observed spread in $v_{m}/\sigma$ for galaxies of a given ellipticity 
is explained as an effect of anisotropy, which changes
the hydrostatic equilibrium.

Observations show that flattening is a common feature of globular 
clusters, which has been known since the early work done by 
Pease \& Shapley (1917). 
Measuring projected ellipticities $e=1-b/a$ of large globular cluster samples 
White \& Shawl (1987) derive a mean $\bar{e}=0.07\pm0.01$ for 99 clusters
in the Milky Way, and Staneva {\em et al.}\ (1996) find $\bar{e}=0.086\pm0.038$
for 173 clusters in M31, with maximum values 0.27 and 0.24 of individual
globulars, respectively.
Kinematical data, i.e. radial velocities of large numbers of cluster members,
reveal that this flattening may indeed
be explained in terms of rotation, and that
the minor axes are nearly coincident with the determined rotation axes
(Meylan \& Mayor 1986). 
Dust obscuration, anisotropy or tidal distortion are able to explain individual
cases of flattening, but can statistically be ruled out as the main mechanism
(White \& Shawl 1987). 

Significant ellipticity variations are found within globular clusters
(e.g., Geyer \& Richtler 1981, Geyer {\em et al.}\ 1983), and these partly also 
coincide with the rotation curves obtained by fits to the radial velocity 
data with some parametrization
specified for the velocity field (Meylan \& Mayor 1986). Moreover, Kontizas
{\em et al.}\ (1990) show that the outer parts of globulars in the Small
Magellanic Cloud are obviously rounder than the parts inside the half mass 
radius and it is likely that their structure differs from that of the galactic
globular clusters because they are younger (in general) and subject to 
different tidal forces.
The importance of age for the interpretation of observed ellipticities
has already been emphasized by Frenk \& Fall (1982), who undertook eye-estimates
of cluster ellipticities in the Milky Way and the Magellanic Clouds, the latter
being slightly larger than the former, which is explained again in terms
of internal globular cluster evolution. This view is supported by studies 
relating Milky Way globular cluster ellipticities to the cluster concentration 
parameter $c=\log(r_{t}/r_{c})$ (White \& Shawl 1987),
where $r_{t}$ is the tidal radius and $r_{c}$ is the core radius, or to the
half mass relaxation time $t_{r_{h}}$ (Davoust \& Prugniel 1990), 
both representing the dynamical age of a globular cluster.
In these two investigations, the average flattening of the dynamically younger 
systems is significantly larger as well, indicating that loss of angular 
momentum, presumably originating from diffusion past the escape velocity
on relaxation time scales, decreases the ellipticity of a cluster.

Indeed, Agekian (1958) suggested a model in which specific angular momentum
is lost due to a relatively large fraction of escaping stars residing 
in the tail of a rotating 
Maxwellian velocity distribution shifted towards the direction of rotation
as compared to the fraction which is counter-rotating.
Considering this effect for every volume element of rotating ellipsoids
he obtained a critical ellipticity of $e= 0.74$ below which the systems
become rounder with time.

More recently, even larger compilations of radial velocity data have been
obtained by different working groups for some globulars. With these data sets
they are able to model the velocity field around the core with nonparametric
methods, and find in the core of 47 Tuc increasing
angular velocities towards the centre (Gebhardt {\em et al.} 1995), 
which is interpreted as a gradually faster
rotating core due to (gravothermal) contraction of an initially slowly rotating
solid body. This would imply that trapping of angular momentum in the 
contracting core wins over angular momentum transport due to viscosity effects.

Analogously to the discussion whether evaporation or gravothermal contraction
drives cluster evolution predominantly, a new question arose due to the 
stability analysis applied to adiabatically confined, rotating cylinders
by Hachisu (1979). He found that, depending on central temperature and
angular velocity, there exist cases where, if angular momentum is removed
from a shell,  gravitational contraction results in an
increase in angular velocity, which leads to a runaway in angular
momentum transport (by viscosity) and central contraction. In analogy to
`gravothermal catastrophe' this effect is called `gravo-gyro catastrophe',
and the physical origin in the latter case is due to a negative specific 
moment of inertia, analogously to negative specific heat in the former case.
In order to study these effects accurately in a cluster-like structure
Akiyama \& Sugimoto (1989) set up N-Body simulations for rotating clusters,
however,
they simulated 1000 particles only. Therefore, the statistical quality
of their data is rather poor and consequently,
the four-phase evolution proposed by them consisting
of {\em i)} violent relaxation,
{\em ii)} gravo-gyro instability, {\em iii)} static evolution, {\em iv)}
gravothermal collapse, seems to be rather suggestive than indicative.

Goodman (1983) performed the first numerical simulations solving the
orbit-averaged Fokker-Planck equation in $(E,J_{z})$-space
applied to rotating stellar systems. His 
main results are that the difference in ellipticities between old galactic
globular clusters and the younger Magellanic clusters can be explained
by relaxation processes, though runs representing appropriate cluster
structure parameters (e.g., smaller King-$W_{0}$-parameters than 9.8) 
were not stated
in his thesis, and that during the self-similar collapse phase the central
angular velocity increases as a small power of central density, which is
explained by linear perturbation analysis performed on the self-similar
solution found by Lynden-Bell and Eggleton (1980).
However, this work suffered from coarse numerical resolution and
from the problem that determinations of asymptotic values 
remained open.

A sequence of rotating King models is established by Longaretti \& Lagoute
(1996), who present an evolutionary path represented by the time
evolution of
structural parameters; their models take into
account evaporation and loss of angular momentum by using
simplified expressions and approximations for the derivation of their
diffusion coefficients. Some of them emerge to be strong constraints to
the general view of evolutionary characteristics of collisional sytems.
In particular, deviations from a rotating King
form of the distribution function may be significant in advanced stages
of cluster evolution.

The present study is aimed to disentangle the effects of evaporation
and angular momentum transport stated above and investigating the
detailed evolutionary scenarios, when initial models are given, thereby
continuing the work begun by Goodman (1983). Moreover, the incorporation
of the dynamical effects of rotation is seen as one of the
stepping stones towards realistic models of globular clusters; 
further steps as they have been studied in the non-rotating case
(post-collapse, mass spectrum, stellar evolution) are the subject of
future papers.
A new self-consistent
Fokker-Planck solver has been
written and is presented together with detailed numerical results.
Section \ref{sec:mod}  gives a description of how the problem is modelled by
the Fokker-Planck equation in an axisymmetric coordinate system together
with the Poisson equation, while a detailed derivation of the 
diffusion coefficients is given in the Appendix. Also, possible limitations
of the method due to a neglect of the third 
integral will be discussed. Section 
\ref{sec:num} provides an
overview of the numerical method, and in Section \ref{sec:res}
numerical results of
the present single mass models are presented. Section \ref{sec:sum}
summarizes the
derived effects of rotation on dynamical evolution, relates it to work
from other authors and gives prospects for future work.

\section[]{Orbit averaged Fokker-Planck equation}
\label{sec:mod}

\subsection[]{Method}

The method described here may be regarded as a continuation of Goodman's
(1983) work and the mathematical formulation used here is very near to
his, but for the reason that it belongs to a part (``Paper III'') of his
Ph.D.-thesis, which was never published elsewhere, a
description will be repeated here. 

In the present study we follow the evolution of the distribution function
$f$ as a function of the energy $E$ and the $z$-component
of the angular momentum $J_{z}$,
both representing velocity variables, with time $t$. We consider $E$ and $J_{z}$
as the only isolating integrals and neglect evident non-ergodicity on the
hypersurface in phase space given by $E$ and $J_{z}$ due to
any third integral. Because the derivation of the diffusion coefficients
necessitates an isolating integral expressed as a function of coordinate
space and
velocity variables, analytical expansions turn out to be unsuitable for
the present
technique and $J^{2}$ as an approximate third integral in the limiting case
of spherical symmetry would be the only choice left to be considered, as it,
e.g., has been used in the approximate 
self consistent dynamical models prepared by
Lupton \& Gunn (1987). But at the current status of numerical resources
we decided to leave the investigations of 3-integral models
for the future.

The Boltzmann equation transformed to the new variables then reads as
\begin{equation}
      \label{eq:bol}
      \frac{\partial f}{\partial t} + \frac{\partial \phi}{\partial t}
                                      \frac{\partial f   }{\partial E}
      = \left( \frac{\partial f}{\partial t} \right)_{\rm enc}   \: ,
\end{equation}
with the potential $\phi$ advanced according to the Poisson equation
\begin{equation}
      \label{eq:poi}
      \nabla^{2} \phi = 4 \pi G n     \: ,
\end{equation}
and the collisional term on the right hand side of Eq. \ref{eq:bol}, 
expressed under the Fokker-Planck assumption of small angle scatterings
\begin{equation}\renewcommand{\arraystretch}{2.0} \begin{array}{lcl}
      \displaystyle \left( \frac{\partial f}{\partial t} 
                  \right)_{\rm enc}  & \displaystyle = &
               \displaystyle   \frac{1}{V}\left( \rule{0ex}{1.5ex}
                      -\frac{\partial}{ \partial E}
                          (<\Delta E>fV) \right. \\
& & \;\;\;\;\;\;\;  \displaystyle
                   -\frac{\partial}{\partial J_{z}} (<\Delta J_{z}>fV)  \\
& & \;\;\;\;\;\;\;  \displaystyle 
         + \frac{1}{2}\frac{\partial^{2}}{\partial E^{2}}(<(\Delta E)^{2}>fV) \\
& & \;\;\;\;\;\;\;  \displaystyle 
             +            \frac{\partial^{2}}{\partial E \partial J_{z}}
                                       (<\Delta E \Delta J_{z}>fV)  \\
& & \left. \;\;\;\;\;\;\; \displaystyle 
      +  \frac{1}{2}\frac{\partial^{2}}{\partial J_{z}^{2}} (<(\Delta J_{z})^{2}
              >fV) \right) \; ,
\end{array}
\end{equation}
where $V$ is the volume element given by $2\pi/\rho$, with $\rho$ being
the radius in cylindrical coordinates.
Because the relaxation time $t_{r}$ is much longer than the dynamical
time $t_{\rm dyn}$ of the system, the Boltzmann equation in our problem
reduces to an equation in which the distribution function only depends
on orbital constants and time. Thus we are allowed to take the orbit average
over that area in the meridional plane which intersects with the hypersurface
in phase space for which $E$ and $J_{z}$ are specified. The condition is
that kinetic energy in the meridional plane be non-negative:
\begin{equation}
      \label{eq:con}
      \frac{1}{2} \left( v_{\rho}^{2} + v_{z}^{2} \right) \gid 0 \: .
\end{equation}
The volume of this hypersurface is given by
\begin{equation}
      p(E,J_{z}) = 4 \pi^2 \int\!\!\!\int\limits_{A(E,J_{z})} d\rho\,dz \: ,
\end{equation}
where the intersection defined by Eq.~(\ref{eq:con}) is given by $A(E,J_{z})$,
and the factor in front of the integral is due to integration over the
third velocity variable, e.g. $\psi = \arctan (v_{\rho}/v_{z})$. Symmetry
about the azimuthal
direction in coordinate space was assumed.

The orbit averaged Boltzmann equation then takes the form
\begin{equation}
      \label{eq:bol2}
      \frac{\partial f}{\partial t} + \frac{1}{p}
                                      \frac{\partial q   }{\partial t}
                                      \frac{\partial f   }{\partial E}
      = \left( \frac{\partial f}{\partial t} \right)_{\rm enc}   \: ,
\end{equation}
in which we introduce the function $q(E,J_{z})$ defined to be
\begin{equation}
      \label{eq:adi}
      q(E,J_{z}) = 2\pi^2 \int\!\!\!\int\limits_{A(E,J_{z})} 
                   \left( v_{\rho}^{2} + v_{z}^{2} \right)     d\rho\,dz \: ;
\end{equation}
note that the integrand is equal to $2(E-\phi)-(J_{z}^2/\rho^2)$ and 
evaluates to zero on the boundary of $A$. Goodman emphasizes that
$q$ is the axially symmetric analogue of the radial action
\begin{equation}
      Q(E,J) = 2 \int^{r_{+}}_{r_{-}} v_{r} dr
\end{equation}
 in a
spherically symmetric potential and comprises actually the average
of the radial action over $J$ with boundaries $J_{z}$ and $J_{max}$,
if the potential is spherically symmetric,
thus representing an average of an adiabatic invariant with respect to
the third integral.
 
Interpreting $q(E,J_{z})$ as an adiabatic invariant itself, equation
\ref{eq:bol2} indicates that, neglecting any encounters,
\begin{equation}
    \left. \frac{\partial f}{\partial t} \right|_{q,J_{z}} = 0 \,   ,
\end{equation} 
from which it follows, that there is a redistribution of energies in the system,
but the 'adiabatic invariant' $q$ and the angular momentum are conserved. 
The desired solution of the Boltzmann equation therefore may be split into
two parts: the first is to calculate the evolution due to stellar collisions
(i.e. small angle scatterings: the Fokker-Planck step) with the
potential held fixed and the second is
to advance the distribution function $f$ due to slow adiabatic changes in the
potential with $f(q,J_{z})=const$.

The Fokker-Planck equation is transformed into flux conservation form in order
to improve conservation of several quantities:
\begin{equation}
     \frac{df}{dt} = \frac{1}{p}\left( - \frac{\partial F_{E}}{\partial E}
                                   - \frac{\partial F_{J_{z}}}{\partial J_{z}}
                                \right)
\end{equation}
with particle flux components
\begin{eqnarray}
     F_{E} & = & -D_{EE}    \frac{\partial f}{\partial E}
                 -D_{EJ_{z}} \frac{\partial f}{\partial J_{z}}
                 -D_{E}     f         \\
     F_{J_{z}} & = & -D_{J_{z}J_{z}} \frac{\partial f}{\partial J_{z}}
                     -D_{J_{z}E}    \frac{\partial f}{\partial E}
                     -D_{J_{z}}    f    \, .
\end{eqnarray}
The orbit averaged flux coefficients $D_{ii}$ in these equations are derived 
from the local diffusion coefficients, which again are found for the
axial symmetric geometry via the prescriptions of Rosenbluth {\em et al.}\ (1957)
involving covariant derivatives instead of the procedure employed by
Goodman (1983) using a non-covariant form. 
The derivation of the flux coefficients is given in
the Appendix. Comparison of the expressions for the diffusion coefficients
found by us and Goodman reveals complete agreement.

The derivation of the diffusion coefficients makes it necessary to specify
a background distribution function by which test stars are scattered. Using
self-consistently the foreground distribution for the background would 
require computational time proportional to $N_{\rho}\times N_{z}\times N_{E}^2
\times N_{J_{z}}^2 $, where the $N_{i}$ are grid sizes in coordinate
and velocity
space, which would reduce the grids employed to inaccurate descriptions
of the current problem; thus, as in all previous applications concerning 
2D Fokker-Planck methods for stellar systems, we set up an appropriate
form of the background distribution. Herein we follow Goodman (1983) giving
a rotating Maxwellian velocity distribution to the background,
\begin{equation}
      \label{eq:max}
      f_{b}(\vec{v}) = \frac{\rho}{(2\pi\sigma^2)^{\frac{3}{2}}}
               \exp(-\frac{(\vec{v}-\Omega \rho \vec{e}_{\varphi})^2 } {2\sigma^2})
\end{equation}
where $\rho$, $\Omega$ and $\sigma$ correspond
to  the zeroth, first and second order 
moments of the distribution function, and are  the density, angular velocity
and one-dimensional
velocity dispersion of the field star distribution, respectively.
Applying this form (Eq.~(\ref{eq:max})) to the background distribution yields
analytical expressions in the diffusion coefficients parametrized locally
by the three moments mentioned above, just thereby reducing the computational
efforts necessary.

\subsection{Initial conditions}
 
For starting configurations rotating King models are utilized.
These are of the form
\begin{equation}
      \label{eq:kin}
   f_{rk}(E,J_{z}) = {\rm const}
  \cdot (e^{-\beta E} -1)\cdot e^{-\beta \Omega_{0} J_{z}}
      \, ,
\end{equation}
where $\beta=1/(m\sigma_{c}^2)$ and the dimensionless angular velocity
$\omega_{0} = \sqrt[]{9/4\pi G\rho_{c}}\cdot
 \Omega_{0}$ are parameters to be specified for each model. In order to
construct a potential-density pair, $\beta$ is related to the
King-parameter (dimensionless potential)
$W_{0}=-\beta m (\phi-\phi_{t})$.
Throughout this paper our different evolutionary runs
are parametrized uniquely by the set of initial conditions given by its
respective pair $(W_{0},\omega_{0})$. In establishing the starting models we 
follow Lupton \& Gunn (1987) and take advantage of an almost
always converging iteration procedure between evaluation of the potential,
brought into the form of a multipole expansion, and the determination
of the density, where different types of Gauss-integrations are performed
and thereby slightly different numerical procedures are employed
(Einsel \& Spurzem 1994).

For simplicity and scaling reasons we choose the gravitational constant $G$,
the initial cluster mass $M_{i}$ and the initial core radius $r_{c_{i}}$
to be one, i.e. to be our system units. The unit of time is expressed as
(Cohn 1979)
\begin{equation}
 \label{eq:t0}
     t_{0} =  \sqrt{\frac{r_{c_{i}}^3}{GM} } \cdot \frac{(GM)^2}{4\pi\Gamma}
              \frac{1}{N}  \, ,
\end{equation}
$\Gamma=4\pi (Gm)^2 \ln \Lambda$, where $N$ is the total number of particles,
$\ln \Lambda$
is the coulomb logarithm and
$m$ is the mean mass of particles (equal to $1/N$ in the present single mass
models). The first factor on the right hand side is simply the
dynamical time of the system and evaluates to one with the units given above,
while the second factor scales out with factors standing in front of
the diffusion coefficients simplifying their expressions appreciably, then.
An expression for the
initial half mass relaxation time is taken from Spitzer \& Hart (1971)
\begin{eqnarray}
     t_{r_{h},i} &  = & 0.138 \sqrt{ \frac{ N \cdot r_{h_{i}}^{3} }
{G\cdot m}}
\frac{1}{\ln \Lambda} \\
          &  = & 32.5  \; , 
\end{eqnarray}
in units of $t_{0}$ (Eq. (\ref{eq:t0})), and we 
assume $\Lambda = 0.4\cdot N$. Due to mass loss the relaxation time may
change by a factor of 2-4 during the runs presented here.

\section{Numerical method}
  \label{sec:num}
 \subsection{Computational grid}

In order to obtain an appropriate grid representation, e.g. a sufficiently
resolved core and rectangular grid, a transformation of the basic
variables was performed. Following Cohn (1979) we take
\begin{equation}
     X(E) = -\ln \left( \frac{E}{2\phi_{c}-E_{0}-E} \right)   \,  ,
\end{equation}
where $E_{0}$ is the energy of a circular orbit at the core radius of
the cluster. Thus, inside the core (roughly $\phi_{c} \lid E \lid E_{0}$)
the relation is nearly linear, while the proportionality
$X\propto -\ln |E|$ in the halo improves
the spacing of the radii of circular orbits with given energies in the direction
of the tidal boundary.
For any specified energy the $z$-component of the angular momentum
is normalized to the maximum angular momentum for that energy, which again
is the angular momentum of the corresponding 
circular orbit as a function of $E$:
\begin{equation}
      Y(J_{z},E) = \frac{J_{z}}{J_{z,0}(E)}  \,  ,
\end{equation}
For each time step $r_{circ}(E)$ and $J_{z,0}(E)$ are determined from the
evolving potential in the equatorial plane by a simple Newton-Raphson scheme.
At the same time the central potential changes, so that the complete
$(X,Y)$-grid is adapted to the new situation. 

\subsection{Fokker-Planck step}

A finite difference scheme is used to integrate the diffusion part of the 
problem and the discretized equation system is solved implicitly with
the help of a sparse matrix method (Henyey 1959) 
borrowed from a gaseous model code
(Spurzem 1994, 1996). A Chang-Cooper scheme is applied to the energy direction
in order to improve conservation characteristics (Chang \& Cooper 1970).

A double-logarithmic coordinate space grid is constructed for the meridional
plane $(\rho,z)$, which extends initially from $10^{-3}$ core radii to somewhat
beyond the tidal radius. During all runs described here it was adjusted
to the contracting core radius $r_c(t)$ by the condition
$\rho_{i=0}=z_{i=0}=0.001\cdot r_{c}(t)$
for the innermost grid point. This was usually done 
only about
4-6 times in each run in order to avoid rounding errors due to the necessary
interpolation procedure involved.

Because of deviations of the local velocity distribution from a rotating 
Maxwellian (Eq.~(\ref{eq:max}))
especially in the outer halo, energy and angular momentum are
not sufficiently conserved if the local foreground values of angular velocity
and velocity dispersion are provided as parameters for the background
distribution function. Instead, these were used as starting values
for an iteration procedure which determines new, consistent background
values for $\omega$ and $\sigma$ under the condition that integrals of
$f_{b}$ times the local first order diffusion coefficients $<\Delta E>$ and
$<\Delta J_{z}>$ (given in the Appendix) over velocity space are zero
(Goodman 1983), expressing the fact that locally no energy and angular 
momentum may be generated or removed just by diffusion. 

\subsection{Vlasov step}

After diffusion has taken place for an appropriate amount of time, the potential
has to be advanced. The density is calculated from the distribution
function according to 
\begin{equation}
     \label{eq:rho}
     n(\rho,z) = \frac{2\pi}{\rho}\int\limits_{\phi_{c}}\limits^{E_{\rm tid}} 
                 \int\limits_{-\rho\sqrt{2E-2\phi}}\limits^{\rho\sqrt{2E-2\phi}}
                 f(E,J_{z})dEdJ_{z}   \,  .
\end{equation}
Given this density distribution, Eq.~(\ref{eq:poi}) can be solved to obtain
the potential. A multipole expansion
is carried out to determine Dirichlet boundary values of $\phi$ beyond the tidal
radius in the meridional plane. Symmetry is assumed about the equatorial
plane so that v.~Neumann boundary conditions
\begin{equation}
    \left. \frac{ \partial \phi}{\partial z} \right|_{z=0}
              =  0  \, , \,
    \left. \frac{ \partial \phi}{\partial \rho} \right|_{\rho=0}
              =  0 
\end{equation}
are applied at the inner boundaries. The system of equations is solved with 
the same sparse matrix method (Henyey {\em et al.} 1959) 
as for the diffusion step, thereby taking account of the even sparser 
matrices occurring due to the Poisson equation.

Having built the potential a new $(E,J_{z})$-grid is created,
the meridional plane is scanned to determine the area $A(E,J_{z})$
and Eq.~(\ref{eq:adi}) is numerically integrated to obtain the adiabatic
invariant $q$. It is required in order to fulfil the conservation
of $f(q,J_{z})$. At first,
the new $(E,J_{z})$-grid is transformed uniquely into a $(q,J_{z})$-grid.
A second order bivariate Taylor expansion is then performed for each 
mesh point in the new grid to derive approximate values from
the known values of $f(q',J_{z}')$ in the old grid. I.e., $f(E,J_{z})$
is allowed to change. This again gives a new density distribution $\rho$
in Eq.~(\ref{eq:rho}) and the procedure is repeated until convergence is
reached. In doing this the firstly determined $f(q,J_{z})$ is retained
as the old grid in order to save accuracy. Nevertheless, this second
order interpolation step turns out to be the
relatively most important source of numerical error in the code.

Goodman (1983) reports a test of the influence of the third integral on the
procedure just stated. He relaxed an initial King model distribution
function (non-rotating) violently in assigning to it a Plummer potential.
The result was a strong dynamical collapse with an increase in $\rho_{c}$
of a factor of 56. The collapse generated a flattening $e=0.055$, which
is attributed to the third integral, because there is no preferential
axis of rotation or anisotropy initially, that could flatten the system.
On the other hand, the test is very crude, because as Goodman himself
states the above procedure is just first order in energy conservation
between initial and final values of $\phi$ and the collapse picture is
not appropriate to be described by adiabatic invariants. Though the
effect of diffusion is to 
wipe out the (here artificially) generated azimuthal anisotropy,  
we consider the flattening of nonrotating, isotropic configurations
during their evolution as a more appropriate measure of the influence
of the third integral.

\subsection{Computation and conservation characteristics}

\begin{table*}
\begin{minipage}{120mm}
 \caption{Initial conditions  of all models presented with $W_{0}=6$.}
 \label{tab:icond}
   \begin{tabular}{l|rrrrrr}
  $\omega_{0}$ &  $T_{\rm rot}/T_{\rm kin}$ & $e_{\rm dyn}(0)$ &
  $r_{\rm tid}/r_{\rm c}(0)$ & $r_{\rm h}/r_{\rm c}(0)$ & $\tau_{\rm rc}(0)$
  & $\tau_{\rm rh}(0)$ \\
     \hline \hline
       0.00   &    0.00    &  -0.001    &  18.72 & 2.70  & 19.24 & 91.88 \\
       0.05   &    0.23    &    0.002   &  18.61 & 2.70  & 19.23 & 91.77 \\
       0.10   &    0.89    &    0.013   &  18.25 & 2.68  & 19.22 & 90.80 \\
       0.20   &    3.38    &    0.051   &  16.83 & 2.66  & 19.20 & 89.71 \\
       0.30   &    7.00    &    0.105   &  14.99 & 2.62  & 19.21 & 87.73 \\
       0.40   &   11.23    &    0.165   &  13.08 & 2.55  & 19.22 & 84.12 \\
       0.50   &   15.61    &    0.224   &  11.46 & 2.48  & 19.27 & 80.49 \\
       0.60   &   19.81    &    0.278   &  9.94  & 2.39  & 19.40 & 76.32 \\
       0.70   &   23.71    &    0.327   &  8.77  & 2.30  & 19.50 & 71.78 \\
       0.80   &   27.18    &    0.368   &  7.69  & 2.20  & 19.71 & 67.37 \\
       0.90   &   30.25    &    0.403   &  6.88  & 2.12  & 19.86 & 63.24 \\
       1.00   &   32.99    &    0.433   &  6.22  & 2.04  & 20.02 & 59.63
   \end{tabular}
 \end{minipage}
\end{table*}

The evolution of the system is followed numerically with time steps
$\Delta t$ chosen to be proportional to the central relaxation time $t_{r_{c}}$
starting with $\Delta t = 0.125\,t_{r_{c}}$ e.g.~in case of a model
with $W_{0}=6$, but the coefficient was increased
from time to time by a factor of $\frac{4}{3}$ in order to have a fractional
increase of the central density between $3-6 \%$. 
In analogy with the computations
done by Cohn (1979) these runs ended with coefficients of about 2. Inbetween
one Vlasov step, i.e. one recomputation of the potential, four
Fokker-Planck steps, i.e. diffusion steps,
were carried out.

In order to stay as close to the King-model distribution functions as possible
we set up a strict tidal cutoff. The tidal boundary is adjusted so as
to ensure that the mean cluster density is conserved throughout the evolution
(see, e.g., Spitzer 1987). The fluxes at the $Y=\pm 1$-boundaries are set
to zero as well as at the $E=\phi_{c}$-boundary. The only open boundary then
is the $E=E_{\rm tid}$-boundary, 
where the condition $f(E_{\rm tid},J_{z})=0$ was fixed.
The first derivative of $f$ with respect to $E$ is non-zero at the
boundary and is evaluated just inside the boundary in order to obtain
accurate escape fluxes.
With grid sizes of $N_{X}=100,N_{Y}=61,N_{\rho}=N_{z}=80$ used here the errors in
mass, energy and angular momentum all accumulated to about $0.4\%$, $0.7\%$ and
$1.7\%$ respectively for a typical model by the time 
the central density increased by about 5-6 orders of magnitude, when 
the runs were stopped. Goodman obtained errors $7.9\%~(12.7\%)$ in mass,
$6.5\%~(2.9\%)$ in energy and $16.5\%~(24.1\%)$ in angular momentum
for runs with $N_{X}=40,N_{Y}=10$ ($N_{X}=20,N_{Y}=10$) 
extending over $2\frac{1}{2}$ orders of magnitude, while Cohn
(1979) reports an error in mass of only $0.04\%$, which may easily be explained
by the application of spherical symmetry to real space in his simulations.

Typical runs for the evolution of one model up to a density increase
of 5 orders of magnitude needed about 60 hours on a 100 Mhz hyperSPARC
workstation (Maths Dept.~in Edinburgh University) or 150 hours on
a 40 Mhz SPARCstation (University of Kiel), respectively. Using a
parallelized version of the code (message passing) on a CRAY T3D 40
hours were needed, when 8 processors were utilized. Although this speedup
is not yet convincing, parallel processing is highly recommended
for multi-mass versions of the present code. 

Although simulations with several choices of $W_{0}$ have been carried out,
the results presented here concentrate on those runs with $W_{0}=6$ leaving
the results with different  $W_{0}$ to the summary section and/or subsequent
papers. Additional features indicated by those do not restrict or alter
the conclusions drawn by using just the results from $W_{0}=6$. The initial
conditions of all models with $W_{0}=6$ are summarized in Table \ref{tab:icond}.
Given are the rotation parameter $\omega_{0}$, the ratio of rotational to total 
kinetic energy, the dynamical ellipticity ({\em cf}.~Eq.~(\ref{eq:edy})), the 
ratios of tidal and half-mass radii to initial core radii, the central
and finally the half-mass relaxation time in system units.

\section{Numerical Results}
\label{sec:res}
\subsection{Evolution of cluster structure}

\begin{figure}
\psfig{file=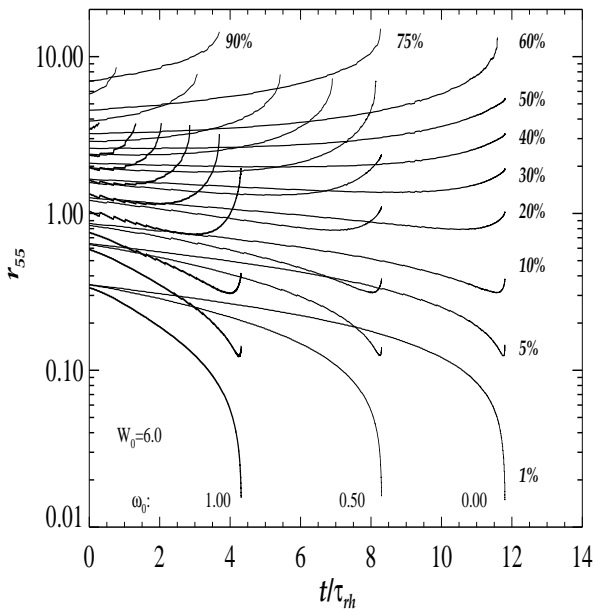,width=120mm,height=85mm,angle=90}
  \caption{Evolution of mass shells (Lagrange radii $r_{55}$) for the model
   $(W_{0},\omega_{0})=(6.0,0.60)$. Shown are the radii for mass 
   columns containing the indicated percentage of total mass
  \label{fig:lag}
   in the direction of the $\theta = 54.74^{\circ}$-angle.
  }
\end{figure}

The evolution of Lagrangian radii with time
for models with $W_{0}=6.0$ and 
$\omega_{0}=0.0,\,0.5,\,1.0$ is shown in Figure \ref{fig:lag}. In flattened
systems the definition of Lagrangian radii is not straightforward. Just only for
that purpose we assume here, since flattening is small, that 
deviations from spherical symmetry are only up to second order in a Legendre
expansion. Thus, we evaluate the Lagrange radii at a zenithal angle, where 
the effects of a probable flattening on the mass columns
are expected to be less important: $P_{2}(\cos \theta)=0$, which gives
$\theta = 54.74^{\circ}$. 
Thereby, we determine $\rho(r,\theta)$ and then compute 
\begin{equation}
  M(r)= \int_{0}^{r} 4\pi \rho(r,54.74^{\circ}) r^2dr \: .
\end{equation}
The tidal radius is determined from
the condition $\phi(\rho,z)=E_{\rm tid}$ and is obtained here for the same angle
as described above. 

The overall structure of evolution of mass shells closely follows what was
given by previous non-rotating, isotropic or anisotropic models (e.g. Aarseth,
H\'enon \& Wielen 1974, Cohn
1979, Giersz \& Heggie 1994a,b, Giersz \& Spurzem 1994) disregarding
the existence of and evaporation through the tidal boundary.
An expansion
of mass shells larger than $50\%$ is observed and an initially smooth -- later
accelerated -- contraction of the inner mass shells can be seen.
The core radius decreases appreciably as well as the core mass.
While no
special sign of a gravo-gyro contraction (i.e. contraction, levelling off and
(gravothermal) contraction again) is detectable, strong mass loss
is evident in this diagram through the loss of complete mass shells. 
The tidal radius, which is not shown directly in the
diagram, decreases in order to maintain
the condition that the mean density in the cluster orbiting around a
parent galaxy is conserved.

\begin{figure}
\psfig{file=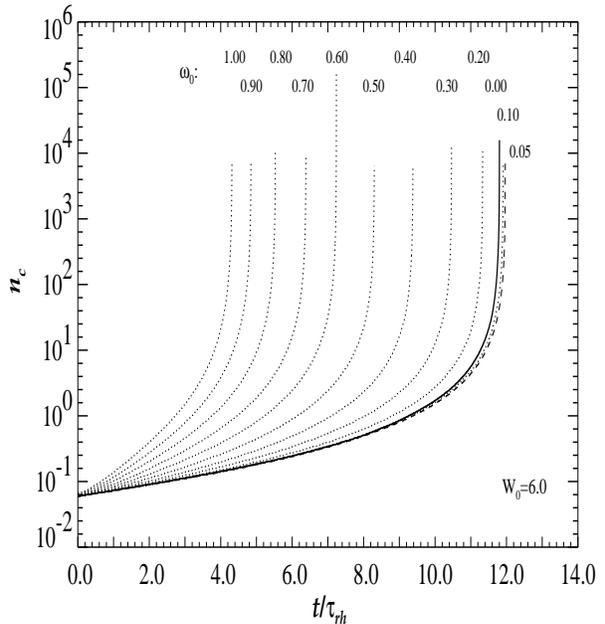,width=120mm,height=85mm,angle=90}
  \caption{Comparison of central density evolution for
   models with same $W_{0}=6.0$, but different initial angular
  \label{fig:den}
   velocity parameter $\omega_{0}=0.0...1.0$. The time is
   given in units of initial half mass relaxation times $t_{r_{h},i}$.
  }
\end{figure}

Figure \ref{fig:den} shows a comparison of the density evolution
of a sequence of models with the same initial $W_{0}=6.0$ but with different 
$\omega_{0}=
0.0 ... 1.0$, respectively. The non-rotating model reaches 
singularity in its core after about $t_{cc}\approx 11.81 t_{r_{h},0}$.
This has to be compared with previously derived collapse time scales,
which were in many cases determined for isolated Plummer models, 
giving $t_{cc}\approx 15.5...17.6 \, t_{r_{h},0}$ 
(Cohn 1979, 1980, Takahashi 1996ab). 
Chernoff \& Weinberg (1990) find times to reach core collapse 
for isolated King models (without stellar evolution) with initial
value $W_{0}=3.0$
($W_{0}=7.0$, $W_{0}=9.0$) of $t_{cc}= 9.6\, t_{r_{h},0}$
($t_{cc}= 10.1 \, t_{r_{h},0}$, $t_{cc}= 2.23 \, t_{r_{h},0}$, respectively).
Note that the latter two values were not published in their paper, but
reported by Quinlan (1996), who himself finds a collapse time for the 
corresponding isotropic ($f(E)$) King model (case of $W_{0}=6.0$)
with tidal mass loss to be $t_{cc}= 12.9\, t_{r_{h},0}$.
Detailed comparison with Quinlan's sequence of King models from 
$W_{0}=1$ up to $W_{0}=12$ reveals a systematic difference of about
$10\,\%$ between our and his results. An increase of the numerical resolution,
i.e.~the grid sizes, in the present models results in slightly larger
collapse times thereby approaching Quinlan's accurate values. 
Regarding the partly large differences between collapse times determined
with distinct methods for (nearly) the same problems
(isotropic or anisotropic models) reported in the literature, we conclude that
our results for the non-rotating models
may be considered as to be in close agreement with previous work on
that field.

On the other hand Fig.~\ref{fig:den} 
gives accurate information about the influence of 
rotation on collapse time. The strongest rotating model with $\omega = 1.0$
gives the smallest $t_{cc} = 4.3t_{r_{h},0}$, implying that rotation accelerates
the collapse. While there is no three- or four-phase structure of 
contraction or collapse phases of different origin (Akiyama \& Sugimoto 1989),
the steeper slope in central density in the higher rotation case indicates 
the existence of a gravogyro contraction, which then transforms into
gravothermal collapse which is more advanced than in the non-rotating case.

\begin{figure}
\psfig{file=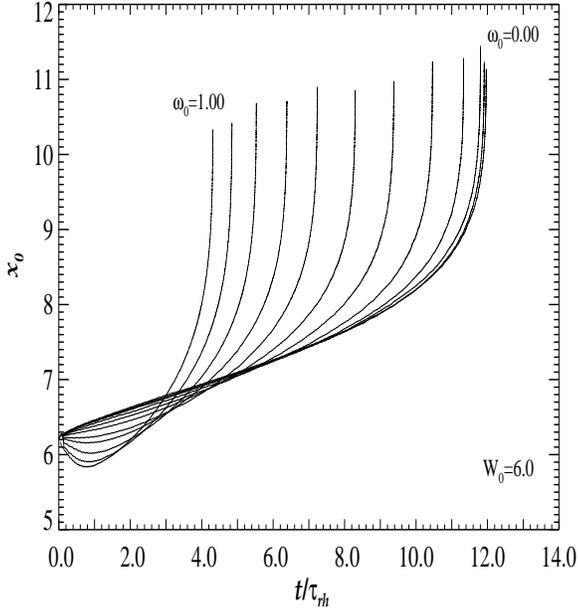,width=120mm,height=85mm,angle=90}
  \caption{Comparison of scaled escape energy evolution for
   models with same $W_{0}=6.0$, but different initial angular
  \label{fig:sca}
   velocity parameter $\omega_{0}=0.0...1.0$. The time is
   again given in units of initial half mass relaxation times $t_{r_{h},i}$.
  }
\end{figure}
 
A similar picture derives from Figure \ref{fig:sca}, where the time
evolution
of the scaled escape energy 
\begin{equation}
      x_{0}= \frac{(\phi_{t}-\phi_{c})}{\sigma_{c}^2}
\end{equation}
($\sigma_{c}$ is the one dimensional central velocity dispersion) is shown.
It is noteworthy that all models show the steep upturn in $x_{0}$, when
they approach the collapse time and they arrive at values of $x_{0}$ between
8.0 and 9.0. This feature (and with this range of values)
has already been described theoretically
by Lynden-Bell \& Wood (1968) and numerically by Cohn (1979), so that we
are in agreement with previous results interpreting this in terms of
gravothermal instability. The effect of rotation can be seen in the earlier
evolution: the model with $\omega=1.0$ is prevented from the usual 
deepening of the potential scaled by the velocity dispersion in the centre
of the cluster and stagnates near its starting value of $x_{0}$. Considering
the stronger increase in central density inferred from Figure \ref{fig:den}
for this model at the same time, the reason for the behaviour
in Fig.~\ref{fig:sca} will be found in a
relatively strong increase in the central velocity dispersion. 
The sources of this effect then subside slowly and a smooth turnover
into the fast collapse is traced, and so this model crosses the paths of the
still gently contracting slow- and non-rotating configurations.

\begin{figure}
\psfig{file=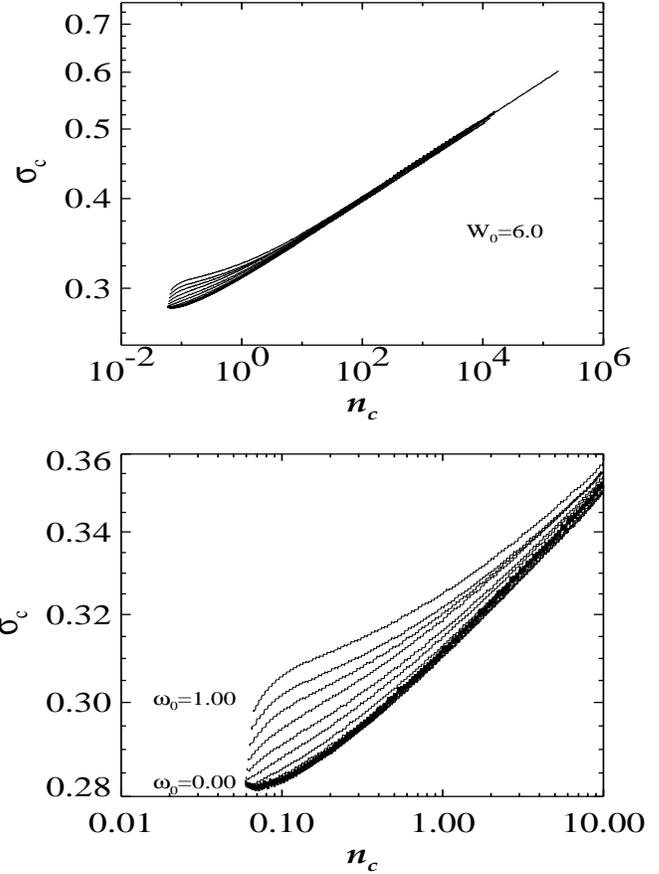,width=90mm,height=120mm,angle=0}
  \caption{Evolution of central velocity dispersion versus central
  \label{fig:dis}
   density for the same models as in Figure 3.
   The curve of the stronger rotating models show a remarkable
   upturn at the beginning (i.e. at low densities).
  }
\end{figure}
 
To clarify these processes we plot in Figure \ref{fig:dis} the central
velocity dispersion against central density. The evolution starts in the
lower left corner of the diagram. While the slowly and non-rotating
models show the typical sign of an initial reluctance to increase
in velocity dispersion as is the case for isotropic and anisotropic
systems (e.g. Cohn 1979), the rapidly rotating model ($\omega_{0}=1.0$)
seems to suffer a significant thermalization and heating of its core. The
evolutionary paths then join again and start the self-similar collapse
phase, which is characterized by the quantity
\begin{equation}
     \gamma = \frac{d\ln \sigma_{c}^{2}}{d\ln \rho_{c}}  \, ,
\end{equation}
which represents twice the slope in the double logarithmic plot ($\sigma_{c}$,
not $\sigma_{c}^{2}$ is plotted). From all models of our 2D-calculations we
derive an average $\gamma = 0.109$ with no dependence of $\gamma$ on the amount of
initial rotation, implying that initial rotation seems to have a minor
influence on collapse characteristics. Cohn's (1979) and
Takahashi's (1996a) 2D-results
show $\gamma = 0.12$ and $0.10$, respectively, and Cohn (1980) obtains 
$\gamma = 0.10$ from his isotropic systems in rough agreement with our
result. Goodman (1983) also derives a value of $0.10$ for $\gamma$ with
his low resolution grids.

\begin{figure}
\psfig{file=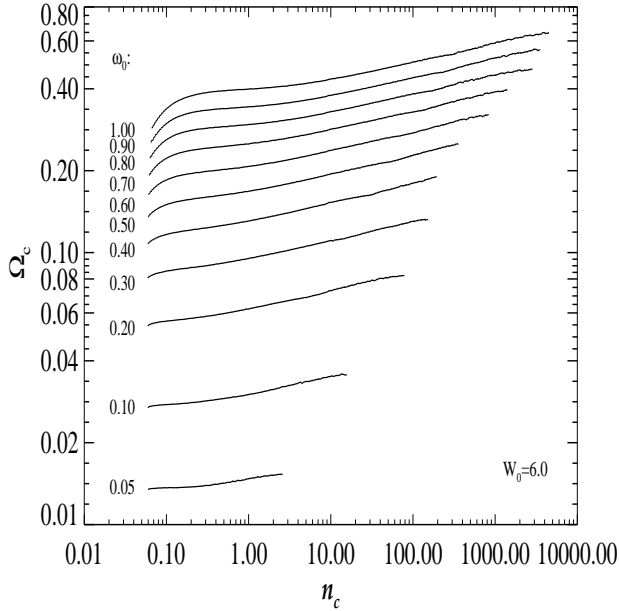,width=120mm,height=85mm,angle=90}
  \caption{Evolution of angular momentum versus central
  \label{fig:ang}
   density for all rotating models with $W_{0}=6.0$.
  }
\end{figure}

The gravo-gyro picture by Hachisu (1979) indicates that the positive variation
of temperature in the core is due to a rise in entropy caused by viscosity
effects, when angular momentum is transferred to disordered motion, and is
thereby
effectively carried outwards (negative specific moment of inertia). In
Fig.~\ref{fig:ang}, a plot of central angular velocity versus central
density is given in the same manner as the $\log(\sigma_{c})-\log(\rho_{c})$
diagram above, but only for the rotating models. There is a pronounced
initial increase in angular velocity for the most rapidly rotating
model,
which after about $2.5~t_{r_{h},i}$ reaches a plateau and finally turns
over into a power law relation of constant slope in this logarithmically
plotted diagram when roughly $4~t_{r_{h},i}$ have passed by. 
The increase in $\omega_{c}$ is connected with the
heating process evident from Figure \ref{fig:dis}. 

The stagnation phase after the initial increase marks the time, when
most of the total angular momentum is diminished (transported into the outer
halo or evaporated) ending the state of negative specific moment of inertia
in the core. The further evolution is now determined by self-similar
evolution in the gravothermal picture, as can be seen if we take a
rough measure of the quantity
\begin{equation}
         \delta = \frac{d \ln \omega_{c}}{d \ln \rho_{c}}  \, .
\end{equation}
Inspection of Figure \ref{fig:ang} reveals $\delta = 0.06 ...0.08$ (with a small
dependence on the amount of initial rotation),
which means, that azimuthal (rotational) velocities in the 
decreasing core
increase in roughly
the same way as any other velocity (inferred from $\gamma/2
\approx 0.055$). Goodman found values $\delta = 0.14$ for grid size
$(N_{X}\times N_{Y})= (20\times 10$) and $\delta = 0.10$ for a 
($40\times 10$)-grid. 

\begin{table}
 \caption{collapse parameters of all simulated models with $W_{0}=6$.}
 \label{tab:w6tab}
   \begin{tabular}{l|rrrrrrr}
  $\omega_{0}$ &  $\tau_{\rm cc}$ &  $\xi$ & $\tau_{\rm rem}$ &
   $\gamma$ & $\delta$
  & $\frac{\Delta M}{\tau_{\rm rh}(0)}$ & $\frac{\Delta J_{\rm z}}
    {\tau_{\rm rh}(0)}$\\
     \hline \hline
       0.00   &  11.804 & 4.69  &   256   &  0.111 &   --  & 2.2  &  --  \\
       0.05   &  11.967 & 5.04  &   239   &  0.113 & 0.045 & 2.2  & 6.3  \\
       0.10   &  11.913 & 4.82  &   247   &  0.107 & 0.071 & 2.3  & 7.1  \\
       0.20   &  11.334 & 4.61  &   261   &  0.112 & 0.079 & 2.9  & 8.7  \\
       0.30   &  10.463 & 4.74  &   251   &  0.106 & 0.077 & 4.0  & 11.6 \\
       0.40   &   9.383 & 4.79  &   249   &  0.108 & 0.075 & 5.5  & 16.0 \\
       0.50   &   8.302 & 5.03  &   240   &  0.114 & 0.074 & 7.3  & 21.4 \\
       0.60   &   7.238 & 4.53  &   263   &  0.107 & 0.066 & 11.4 & 32.3 \\
       0.70   &   6.387 & 4.84  &   245   &  0.111 & 0.068 & 14.1 & 39.2 \\
       0.80   &   5.532 & 4.70  &   252   &  0.104 & 0.061 & 20.4 & 56.7 \\
       0.90   &   4.850 & 5.04  &   237   &  0.109 & 0.066 & 24.5 & 71.1 \\
       1.00   &   4.318 & 4.89  &   244   &  0.108 & 0.065 & 29.9 & 92.5
   \end{tabular}
\end{table}
Several useful quantities derived from the models presented here are shown in
Tab.~\ref{tab:w6tab}. In column (1) the initial angular velocity in system
units is given ($W_{0}=6$), col.~(2) gives the collapse time, col.~(3) the
collapse rate, col.~(4) the number of current central relaxation times, 
$\tau_{\rm rem}$, until
complete collapse in the self-similar evolution phase, col.~(5) and (6) the
exponents used in the corresponding equations of state of the core, while
col.~(7) and (8) state the percentage of initial mass and angular momentum loss
per half mass relaxation time. Note, that the non-rotating model gives
mass loss rates in full agreement with those values reported in the literature
(e.g.~Spitzer 1987), with strongly increasing values towards models of
higher rotation.
Unfortunately, the well known value of 320 for $\tau_{\rm rem}$ (Cohn 1980) is 
underestimated by our models, which indicates that these suffer from
worsening numerical resolution, when the collapse proceeds further and
further.

\vfill

\subsection{Evolution of cluster shapes}

\begin{figure}
\psfig{file=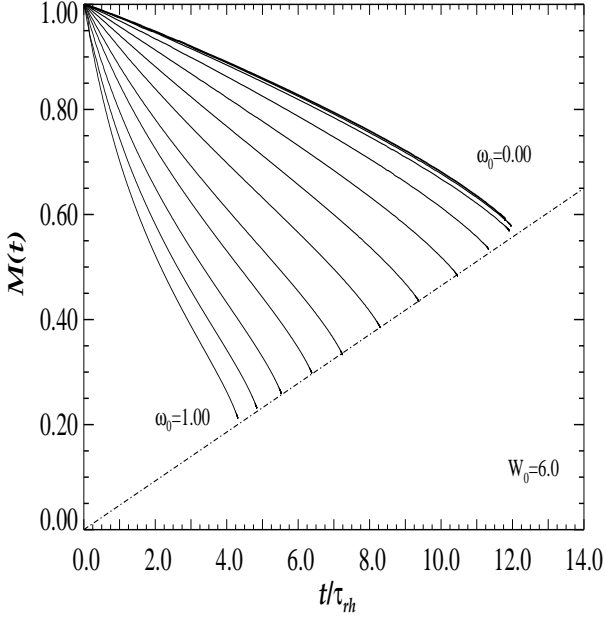,width=120mm,height=85mm,angle=90}
  \caption{Evolution of total mass retained in the cluster
  \label{fig:mas}
   models with $W_{0}=6.0$.
  }
\end{figure}
 
As has already been shown in Figure \ref{fig:lag}, our models -- subject
to the strict tidal boundary formulation given above -- suffer strong
mass loss, which can also be seen from Figure \ref{fig:mas}. Here, a comparison
of preserved mass as a function of time
between the models with same $W_{0}=6.0$ but different rotation 
($\omega_{0,i}=0.0...1.0$) is given. 
In general this diagram reveals information about evaporation 
of angular momentum from the halo, which 
drives cluster evolution with respect to rotation.
As expected the fastest rotating
model shows strongest mass loss indicated by the slope of the curves, and
within its evolution an even steeper mass loss can be found in the
initial two half mass relaxation times, which were associated with the
gravogyro phase of the system.

\begin{figure}
\psfig{file=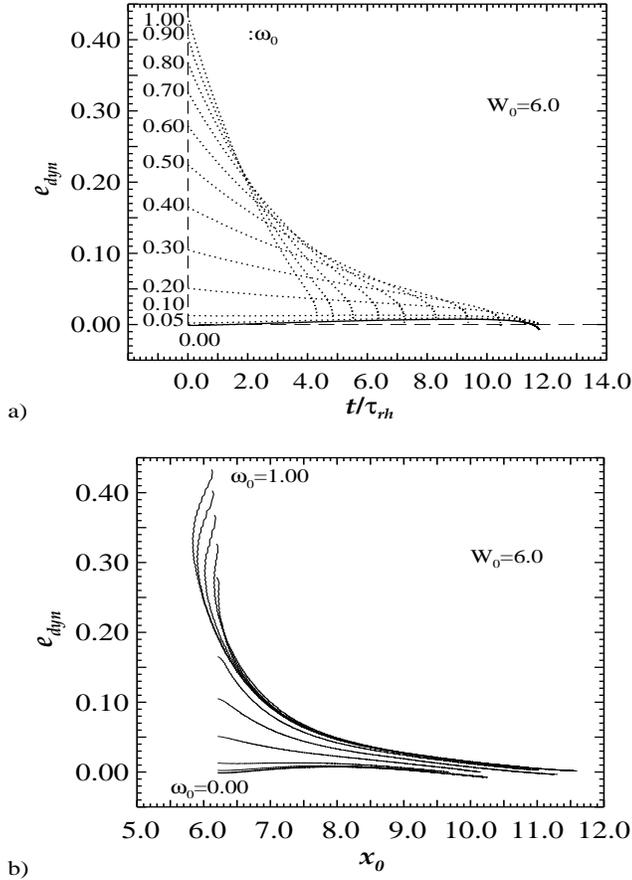,width=85mm,height=120mm,angle=0}
  \caption{Evolution of dynamical ellipticity $e_{\rm dyn}$ as defined
   by Goodman (1983) (Eq.~(27)
  \label{fig:edy}
   see text for explanation) for 
   all models with $W_{0}=6.0$. Fig.~a) gives the evolution versus time
   measured in units of the half mass relaxation time, while Fig.~b)
   shows the evolution versus the scaled escape energy $x_{0}$.
  }
\end{figure}

By transforming formulae derived by Binney (1978), who applies the
tensor-virial theorem to a system composed of
similarly situated ellipsoids, Goodman (1983) defines the
so-called dynamical ellipticity $e_{\rm dyn}$, which then is given by
\begin{equation}
    \label{eq:edy}
     \frac{2T_{rot}+3T_{\sigma_{\phi}}-T_{\sigma}}{T_{\sigma}-T_{\sigma_{\phi}}}
     =  \frac{(1+2s^{2})\arccos s ~-~ 3s\sqrt{1-s^{2}} }
             {s\sqrt{1-s^{2}}~-~s^{2}\arccos s} \, ,
\end{equation}
where $s\equiv b/a = 1-e_{\rm dyn}$ is the axis ratio of the ellipsoids,
which are assumed to be oblate spheroids (i.e. not triaxial),
$T_{rot}$ is the rotational energy, $T_{\sigma_{\phi}}$ is the energy
contained in the azimuthal component of the velocity dispersion and
$T_{\sigma}$ is the energy associated with all components of the
velocity dispersion. For sufficiently small $e_{\rm dyn}$ this relation
may be expressed as
\begin{equation}
     \frac{2T_{rot}+3T_{\sigma_{\phi}}-T_{\sigma}}{T_{\sigma}-T_{\sigma_{\phi}}}
     \approx \frac{8}{5} e_{\rm dyn} \, .
\end{equation}
The evolution with time of $e_{\rm dyn}$ (calculated from Eq.~(\ref{eq:edy}))
is plotted in Figure \ref{fig:edy} for the three models
with $W_{0}=6.0$. A steep decrease in ellipticity can be seen for
the initially most 
strongly rotating model and the final states of all models lack 
significant flattening.
Due to less effective mass loss, the angular momentum transport beyond
the tidal boundary is smaller in the more moderately rotating models, 
so that the respective curves are able to cross
each other.

Spherical symmetry is not exactly preserved for the non-rotating
model (see Fig.~7, curve $\omega_0=0$), 
which can be explained by the creation of anisotropy
in its halo, when azimuthal pressure is compared with that within the
meridional plane. Because there is a preferential direction due to
the chosen coordinate system, i.~e.~the $z$-direction,
no anisotropy may develop in planes tangential to the poles of the system.
Thus, just as radial anisotropy develops in general spherical
systems (e.g., Cohn 1979, Louis \& Spurzem 1991), it evolves in the
equatorial plane of our model configurations, but not along the $z$-axis
with $\rho\ll r_{c}$, which is caused by the neglect of the third integral.
Moreover the present formulation of the tidal boundary
produces an anisotropy profile, which shows
the usual rise towards the halo starting from the isotropic core, but then,
as it approaches the tidal boundary, falls below zero, because stars on radial
orbits are evaporated more effectively than stars on tangential orbits: the
former easily gain energy in core passages, so that those orbits 
are deplenished.
On the other side stars visiting regions near the poles are all coupled
to the core, because no third integral specified 
prevents them from doing that (e.g.,
$J^{2}$ in the spherical symmetric case, if it is non-zero for the orbit
given).

Neglect of the third integral also has effects on orbits when slow
potential changes take place. Consider circular orbits in the meridional plane
and equatorial plane and assume a currently spherical potential. A gentle
contraction of the system will retain a circular orbit in the equatorial plane,
since it is represented accurately by its $(E,J_{z})$-pair (it will just be
shifted to that energy, where $J_{z,0}(E)=J_{z}^{old}$ for angular momentum
has to be conserved), while the former orbit is not distinguished by our 
coordinate systems from radial orbits of the same energy. Therefore, all
orbits with $J_{z}=0$ are shifted to new energies (via the condition of
adiabatic invariant conservation) corresponding to the ensemble mean with
respect to third integral, and this may not be the same energy in the
meridional circular orbit case as that of the equatorial circular orbit.
In particular, the probability of retaining a circular orbit is not
necessarily one. 

Nonetheless, inspection of Figure \ref{fig:edy} indicates, that the
ellipticity of the non-rotating model 
rises due to the neglect of the third integral
only to values $e_{\rm dyn}\approx 0.01$, well below the uncertainties inferred from
observations.

\begin{figure}
\psfig{file=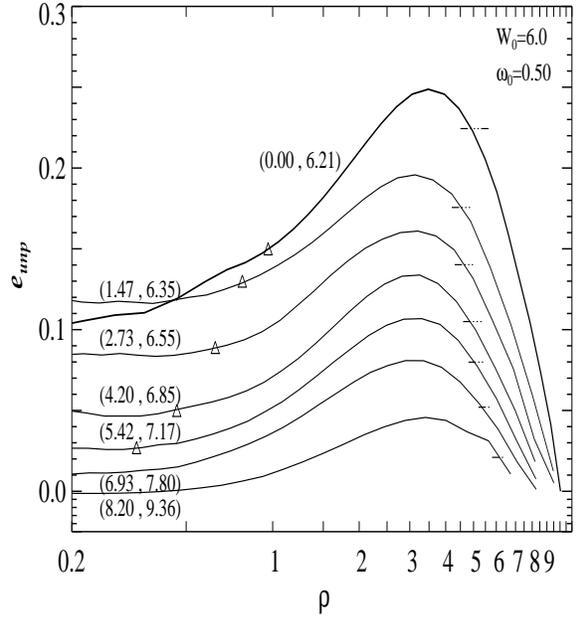,width=120mm,height=85mm,angle=90}
  \caption{profile of unprojected ellipticity for model with 
  \label{fig:ell}
    $W_{0}=6.0$ and $\omega_{0}=0.50$ for several evolutionary times
    as indicated. The pairs in brackets give $x_{0}$ and $t/t_{\rm rh}$, 
    respectively. The horizontal lines on the right side indicate
    the corresponding dynamical ellipticities.
  }
\end{figure}

\begin{figure}
\psfig{file=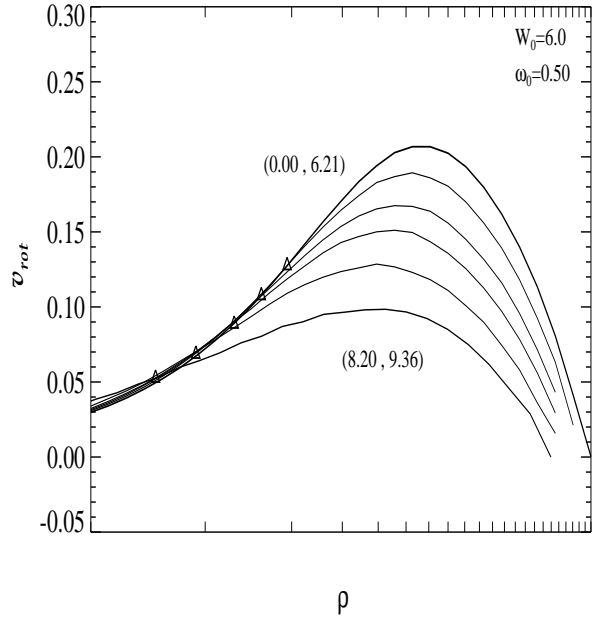,width=120mm,height=85mm,angle=90}
  \caption{profile of rotational velocities $v_{\rm rot}$ for model with
  \label{fig:vrot}
    $W_{0}=6.0$ and $\omega_{0}=0.50$ for several evolutionary times
    as indicated. The pairs in brackets give $x_{0}$ and $t/t_{\rm rh}$,
    respectively. The sequence of curves starts with the pair $(0.00,6.21)$
    at the top. The actual core radii are denoted by the triangles put
    on the curves, respectively. The core radius of the last model plotted
    is situated left from the figure area.
  }
\end{figure}

\begin{figure}
\psfig{file=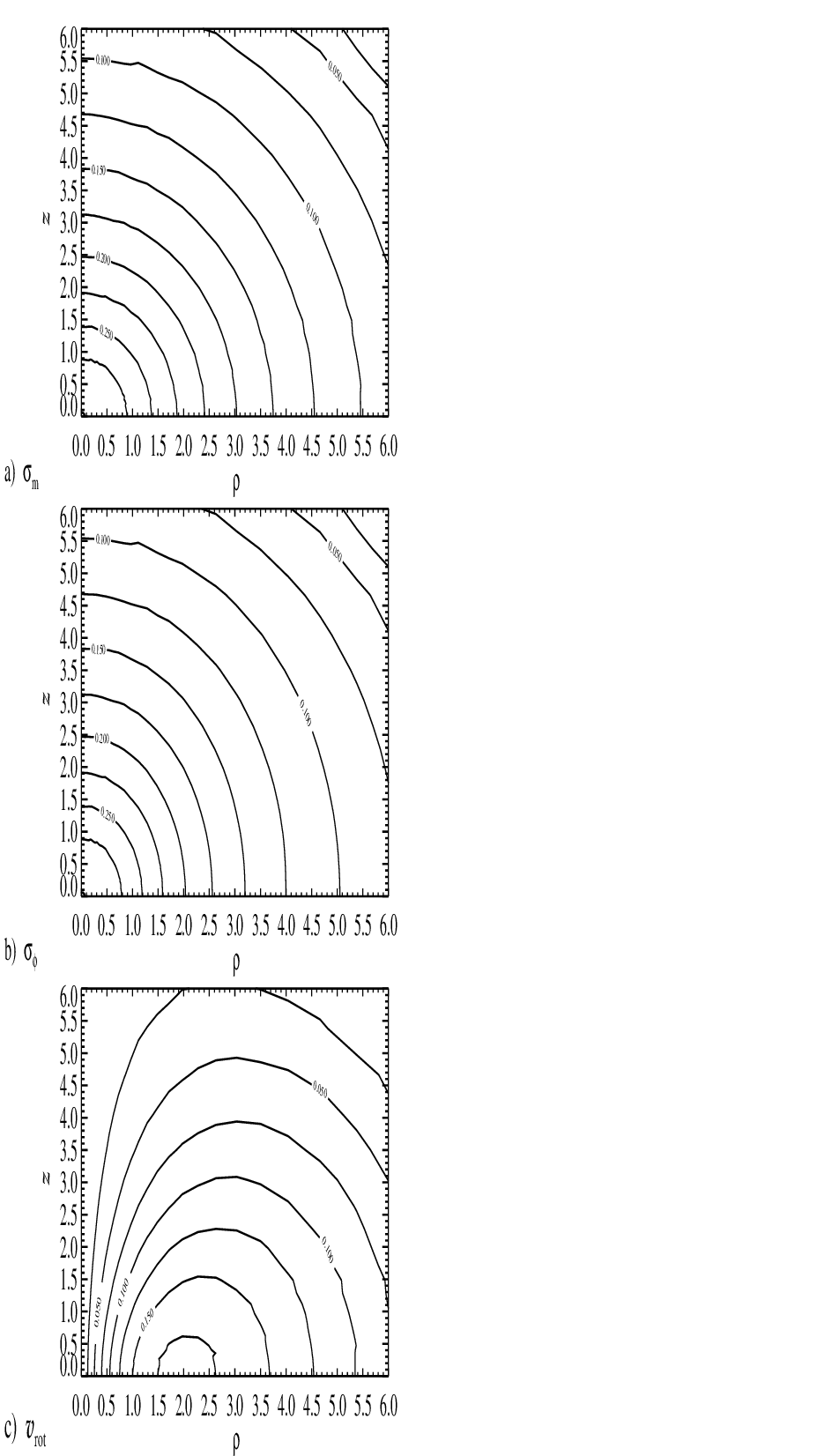,width=157mm,height=220mm,angle=0}
  \caption{Contour maps a) of velocity dispersion in the meridional plane
   b) of the velocity dispersion in azimuthal direction and c) the
   rotational velocity for a model with $W_{0}=6$, $\omega_{0}=0.6$ and
   $t=\tau_{rh_{0}}$.
  \label{fig:oCen}
  }
\end{figure}

In Figure \ref{fig:ell} the variation of unprojected ellipticities
$e=1-b/a$ with radius is shown for the model with $\omega_{0}=0.50$
for several evolutionary times. 
The current core radii are indicated by
triangles on the curves, respectively. 
The ellipticities of the core are evidently smaller as is the case already
for the starting models. The ellipticity peaks at about the half mass radius
where it is appreciably larger than in the core but with a nearly
constant offset
($\Delta e \approx 0.07$ in this case).
This indicates, that we indeed expect a significant ellipticity variation
within globular clusters as has been stated by Geyer {\em et al.}\ (1983).
It is remarkable that the contraction or decrease of the core is not reflected
in this picture. The horizontal lines in this Figure give for comparison the
respective dynamical ellipticities $e_{\rm dyn}$ for that time step. It seems, 
that this quantity represents very well a global mean unprojected ellipticity.

Strong evidence for rotation in globular clusters had been derived
from the observed coincidence of ellipticity and rotational velocity
profiles, both being scaled arbitrarily (Meylan \& Mayor, 1986). 
In Figure \ref{fig:vrot} we show for comparison
for the same model and time steps as in Fig.~\ref{fig:ell}
the corresponding profiles of  rotational velocity.
Prior to gravothermal instability a coincidence of the profiles just
stated may indeed be watched, while for the rotational velocity curve
corresponding to the gravothermal collapse phase 

Recently, maps of rotational velocities and azimuthal as well as
meridional velocity dispersions have been constructed from non-parametric
fitting of large samples of kinematical data for $\omega$~Cen (Merritt
{\em et al.} 1997). Disregarding for the moment the simplicity of the
present models (i.e.~single mass models, no stellar evolution, etc.) we
chose a model from our simulations, which roughly reproduces the
ellipticity and concentration parameter (0.12 and 1.36, respectively) 
derived for $\omega$~Cen from observations. The velocity
maps obtained ($W_{0}=6$, $\omega_{0}=0.6$, $t=3.77~\tau_{rh_{0}}$)
are shown in Fig.~\ref{fig:oCen}.
These maps agree with those from Merritt
{\em et al.} (1997) in the morphological structure,
e.g.~the oblate isovelocity contours in the case of $\sigma _{\phi}$ or
the torus-like contours when the rotational velocity map is considered.
Note, that the axes are measured in units of initial core radii and that
the current core radius has already decreased to about $0.44~r_{c_{i}}$.

\section{Summary and discussion}
\label{sec:sum}

We have performed 2D-Fokker-Planck simulations modelling the evolution
of rotating stellar systems. The main results can be summarized as follows:
large amounts of initial rotation drive the system into a phase of strong
mass loss while it contracts slightly. The core is rotating even
faster than before although angular momentum is transported outwards.
At the same time the core is heating. Given these features we associate
this phase with the gravo-gyro phase found by Hachisu (1979). The total 
collapse
time is shortened appreciably by this effect, but the increase in central
angular momentum levels off after about $2-3~t_{r_{h},i}$ indicating that
the source of this `catastrophe' ceases, i.e.~it is not really a catastophe. 
Finally, the central angular velocity increases again, but with a rather 
small power of the central density -- nearly
the same power as for the central velocity dispersion
during self-similar contraction. 
We propose to search for a self-similar solution of rotating cluster models
in the future, which should agree with our findings.

Due to the non-locality of relaxation, the processes of mass loss,
angular momentum transport and gravo-gyro and gravothermal `catastrophe' in our
models are not easily disentangled and as well are
not entirely the same as e.g.~in rotating gas spheroids. Additionally,
even in that
case no theoretical formulation of the problem has been found, yet. The
only theoretical investigations undertaken 
so far concern rotating gas cylinders (Inagaki \& Hachisu 1978, Hachisu 1979)
and rotating gas disks (Hachisu 1982). But the mechanisms taking place
in our comparatively complicated structure of a slightly flattened
spheroid may be identified by analogy with an idealized transition
between the cylinder and disk configurations of Hachisu (1982). The
mechanism of gravo-gyro `catastrophe' had been introduced by Hachisu
(1979) for rotating gas cylinders, but these models do not comprise
gravothermal `catastrophe' for reasons concerning the curvature
of the metric configuration (Hachisu 1982). On the other hand the 
investigation of interactions
between gravogyro and gravothermal processes may be accomplished by 
considering the disk configuration in which both processes are active.
There, it is shown that those terms in the tensor, which describes
the hydrostatic readjustment of the system to a variation, that denote
a coupling of quantities (e.g.~angular momentum with temperature) play an
important role for further evolution, thereby, e.g., adjusting the gravogyro
to the gravothermal contraction. Then, the time scale
of evolution will be given just by the efficiency of heat transport.

The time needed to reach collapse for each model is consistent with
the individual mass loss rate (see Fig.~(\ref{fig:mloss}), which itself 
is connected with the amount
of angular momentum transport due to viscosity. 
The larger the initial amount of rotation the stronger the mass loss (see
Fig.~\ref{fig:mloss}).
Contrary to the simple models of Agekian (1958)
stellar escapers from the system reach their escape energy in encounters
mostly inside or near the core.

Due to both features -- the total time to reach collapse depends on
the amount of mass loss induced by viscosity effects and the solution for
the gravothermal collapse is only moderately
modified by gravogyro effects -- it may be concluded, that 
the gravogyro `catastrophe' is coupled to the gravothermal contraction
and therefore does not evolve on time scales of angular momentum transport
(viscosity coefficient) but rather on time scales of heat transport. This
is clearly indicated by the lack of an early, separate central contraction
phase of Lagrangian radii (see Fig.~(\ref{fig:lagr})) as it seemed to be 
implied by the N-body simulations of Akiyama \& Sugimoto (1989).
Note, however, that they did not include a tidal boundary.
On the other hand gravogyro effects influence cluster evolution indirectly
by increasing the mass loss activity thereby reducing the total collapse
time.

One of the main results of this work 
is that a cluster system presently comprised of
nearly spherical clusters may have had a distribution of cluster
ellipticities up to strongly flattened, rotating stellar systems.
This raises question for the origin of globular clusters: the picture
presented here would be fully consistent with a formation from
(obviously significantly rotating) giant molecular cloud like structures,
when comparing their mass and angular momentum content (Akiyama \& Sugimoto
1989).  
Other mechanisms, which are still not in contradiction to our results,
incorporate the formation of rotating globular clusters from
previously existing binary clusters,
which suffer a synchronization
instability leading to a merger (Sugimoto \& Makino 1989). It is found
that employing this mechanism a maximum flattening $e\approx 0.3$ 
of the globular clusters can be explained. The advantage of this model is
that it explains very well several features of the Magellanic Clouds
globular cluster system. The stronger flattening of Magellanic Cloud
clusters as compared to those of the galaxy is in the light of our
simulations understandable as a consequence of their dynamical age.
Another possible mechanism to form binary
clusters is collapse from a shell (Theis 1996). The formation of (single)
clusters from collapsing shells has been proposed by
Brown {\em et al.}\ (1991, 1995).

The rotation curves decrease with time, as do the ellipticity profiles,
but the relative shapes of the latter
do not vary much. The maximum values of rotational
velocity and ellipticity remain at about the half mass radius throughout
the evolution.
In accord with our results Gebhardt {\em et al.}\ (1995) observe in 47Tuc
by using non-parametric techniques an increase in the rotational
velocity towards the half-mass radius and nearly solid body rotation
inside the core.
However, using the rotational velocity derived from integrated light
(Gebhardt {\em et al.}\ 1994) they obtain an increasing angular velocity for the
region inside half a core radius.  
Moreover, 47 Tuc reveals an already advanced stage of evolution, so that
the results presented here indicate that a maximum in the velocity curve
occurs at tens of current core radii (preliminary modelling gave about
$20r_{c}(t)$ for a 47 Tuc - like model cluster). While the first argument
may be reconciled with the results presented here (solid body
rotation in the core and a rotation velocity peak at about the half-mass
radius, which is roughly 10 core radii in the case of 47Tuc), 
when the bad statistics
of the data beyond $\approx 2 r_{c}$ in the observations are considered,
the rise in angular velocity in the innermost central parts of the core,
which are indicated by those observations may not be explained by
our actual single mass models. There are several possible reasons, why
our models may not yet be appropiate for 47Tuc: i) in 47Tuc as in any
other cluster near core bounce there is significant activity of hard binaries
going on, whose reaction products have been observed 
(Meylan {\em et al.} 1991); the
present models do not yet incorporate binary energy generation; ii)
the cluster mass function is observed to vary with radius  and time both
in spherical models and observations. Models including mass segregation
effects may differ significantly from the single-mass case, as it is 
known for non-rotating clusters. For example it may be speculated that in deep
core collapse high mass stars, which quickly segregate towards the
centre, cannot loose angular momentum as efficiently as the stars
in our simple models with equal masses do.
iii) Finally, due to its very peculiar
structure 47Tuc could be a non-axisymmetric cluster due to a recent
encounter with the galactic disk, bulge or other cluster.

Future work, which will be published in the next papers of this series,
consists of the incorporation of an energy source due to formation and
hardening of three-body binaries (Bettwieser \& Sugimoto 1984,
Heggie \& Ramamani 1989, Lee, Fahlman \& Richer 1991) in order to
proceed into the post-collapse phase, 
and the yet completely uninvestigated influence of differently rotating
mass groups.
Mass segregation would presumably alter the efficiency
of angular momentum transport significantly.
Such models will allow more detailed modelling of observed clusters, such
as 47 Tuc or others.

Also it is important to consider the effects of different
tidal boundary conditions and the dynamical influence
of mass loss due to stellar evolution, 
both of which can alter the cluster structure considerably.
As for the treatment of the tidal boundary it affects e.g.
the anisotropy profile near the boundary.
Drukier's (1995) boundary formulation with
delayed mass loss beyond the tidal boundary, respectively,
are worth
trying out in this context.
Finally, since the gravogyro contraction
seems to take place in the very early evolution of the cluster, different
initial conditions also may have important bearings on the rotational
structure of the system.

We stress the necessity to perform N-body simulations of rotating
systems in order to indicate the correctness of the assumptions
made in the present Fokker-Planck models, especially to see the
effects due to the neglect of the third integral.

\section*{Acknowledgments}

% C.E. wishes
We wish 
to thank D.~C.~Heggie for kind hospitality during
% his
research visits in Edinburgh and for many interesting discussions as
well as helpful comments and suggestions. Moreover, it is a pleasure
% for C.E. 
for both authors
to thank K.~Takahashi and S.~Inagaki for several
fruitful discussions and kind hospitality during
% his
visits in Japan. The results presented are
obtained in partial fulfillment of the Ph.D. requirements (C.E.) at
Kiel University. This work was supported by DFG-grants Sp345/3-1 and /3-2
including a supplementary SPARCstation. C.E.~is also grateful for support
under the German-Japanese DFG-JSPS program (S.~White, K.~Nomoto) grant no.~
446 JAP 113/109/0. Support for computing and visiting
was provided by EPCC-staff, Edinburgh, in the frame of the TRACS scheme.

\appendix

\section[]{Derivation of the flux coefficients}

When deriving diffusion coefficients several ways may be chosen for
doing it. We decided to follow the approach of Rosenbluth, MacDonald \&
Judd (1957) involving covariant derivatives of tensorial objects.
At first, the coordinate systems will be introduced. While we
have cylindrical coordinates in coordinate space ($\rho , z, \varphi $), 
the following
 symmetry is applied to velocity space: $q^{1}=v=(v_{\rho}^2+v_{\varphi}^2
+v_{z}^2)^{\frac{1}{2}}$; $q^{2}=\psi=\arctan (v_{\rho}/v_{z})$;
$q^{3}=v_{\varphi}$, where $(v_{\rho},v_{\varphi},v_{z})$ are local
cartesian velocity coordinates. 
The corresponding metric tensor $(a^{\mu \nu})$ then
reads as
\begin{equation} 
  a^{\mu \nu} = \left( \begin{array}{ccc}
                       1    &    0    &   \frac{v_{\varphi}}{v}  \\
                       0    & \frac{1}{(v_{\rho}^2 + v_{z}^2)} & 0 \\
                       \frac{v_{\varphi}}{v} & 0 & 1
                       \end{array}  \right)
\end{equation}  
with the volume element $a:=\det(a_{\mu \nu})=1/\det(a^{\mu \nu})=v^2$.
The tensorial form of the Fokker-Planck equation may generally be written as
\begin{equation}
  \frac{1}{\Gamma_a} \frac{\partial f_a}{\partial t} = -(fT^{\mu}_a),_{\mu}
                                      +\frac{1}{2}(fS^{\mu \nu}_a),_{\mu \nu},\:
\end{equation}
where the commas denote covariant derivatives, and the subscript $a$ indicates
particle species. The factor $\Gamma_a=4\pi Gm_a^2 \ln \Lambda$ contains
the usual Coulomb logarythm. The diffusion coefficients of Cartesian
coordinate systems may then be expressed
as tensorial objects
\begin{equation} \label{eq:DV2}
       \frac{1}{\Gamma_a}<\Delta v^{\mu}>_a = T_a^{\mu} = 
             a^{\mu \nu} (h_a),_{\nu}
\end{equation}
\begin{equation}
       \frac{1}{\Gamma_a}<\Delta v^{\mu} \Delta v^{\nu}>_a =
             S^{\mu \nu} =  a^{\mu \omega} a^{\nu \tau} (g),_{\omega \tau}.
\end{equation}
The functions $h$ and $g$ are the so called Rosenbluth potentials:
\begin{equation}
h_a(\vec{v})=\Sigma_b \frac{m_a + m_b}{m_b}\int d\vec{v}'_f f_b(\vec{v}'_f)
                              \frac{1}{\left| \vec{v}-\vec{v}'_f\right|}
\end{equation}
\begin{equation}
g(\vec{v}) = \Sigma_b \int d\vec{v}'_f f_b(\vec{v}'_f) \left| 
                                           \vec{v}-\vec{v}'_f\right|
\end{equation}

After some lengthy calculations involving Christoffel symbols we arrive at
the following expressions for the tensors given above (note, that symmetry
is assumed about $\psi$):
\begin{equation} \label{eq:TENB}
T_a^1 = \frac{\partial h}{\partial v} + \frac{v_{\varphi}}{v}
                 \frac{\partial h}{\partial v_{\varphi}}
\end{equation}
\begin{equation}
T_a^2 = 0
\end{equation}
\begin{equation}
T_a^3 = \frac{v_{\varphi}}{v} \frac{\partial h}{\partial v}
                 + \frac{\partial h}{\partial v_{\varphi}}
\end{equation}
\begin{equation}
S^{11} = \frac{\partial^2 g}{\partial v^2} + 2\frac{v_{\varphi}}{v}
         \frac{\partial^2 g}{\partial v \partial v_{\varphi}}
         + \frac{v_{\varphi}^2}{v^2} \frac{\partial^2 g}{\partial v_{\varphi}^2}
\end{equation}
\begin{equation}
S^{12} = S^{21} = 0
\end{equation}
\begin{equation}
S^{13} = S^{31} = \frac{v_{\varphi}}{v} \frac{\partial^2 g}{\partial v^2}
         + (1 + \frac{v_{\varphi}^2}{v^2})
           \frac{\partial^2 g}{\partial v \partial v_{\varphi}}
         + \frac{v_{\varphi}}{v} \frac{\partial^2 g}{\partial v_{\varphi}^2}
\end{equation}
\begin{equation}
S^{22} = \frac{1}{v(v^2-v_{\varphi}^2)} \frac{\partial g}{\partial v}
\end{equation}
\begin{equation}
S^{23} = S^{32} = 0
\end{equation}
\begin{equation} \label{eq:TENE}
S^{33} = \frac{v_{\varphi}^2}{v^2} \frac{\partial^2 g}{\partial v^2}
         + \frac{\partial^2 g}{\partial v_{\varphi}^2}
         + \frac{(v^2-v_{\varphi}^2)}{v^3} \frac{\partial g}{\partial v}
         + 2\frac{v_{\varphi}}{v} \frac{\partial^2 g}{\partial v_{\varphi}^2}
\end{equation}

Thus, employing the relations 
\begin{equation}
(fT^{\mu}_a),_{\mu} = \sqrt{a}^{-1}\frac{\partial}{\partial q^{\mu}}
       \left( \sqrt{a}fT^{\mu}_a) \right)
\end{equation}
\begin{equation} \begin{array}{ccl}
(fS^{\mu \nu}_a),_{\mu \nu} & = & \sqrt{a}^{-1}
                   \frac{\partial^2}{\partial q^{\mu}
                  \partial q^{\nu}} \left( \sqrt{a}fS^{\mu \nu}\right) \\
& & \\
& & + \; 
    \sqrt{a}^{-1} \frac{\partial}{\partial q^{\nu}}\left(\sqrt{a}\Gamma^{\nu}
        _{\omega \mu} fS^{\mu \omega} \right)
\end{array}   \end{equation}
with the $\Gamma$ symbol denoting a Christoffel symbol of the second kind, here,
the Fokker-Planck equation consists of terms
\begin{equation} \label{eq:FT}
\begin{array}{llr}
\displaystyle (fT^{\mu}_a),_{\mu} &\displaystyle  = & \displaystyle \frac{1}{v}
                      \left( \frac{\partial}{\partial v}
                            \left( vf \frac{\partial h}{\partial v}
                 +   v_{\varphi}f \frac{\partial h}{\partial v_{\varphi}}
           \right) \:
 \right. \\
 & & \\
          & & \left.   \D
                      +  \frac{\partial}{\partial v_{\varphi}}
                            \left( v_{\varphi}f \frac{\partial h}{\partial v}
                      +     vf \frac{\partial h}{\partial v_{\varphi}} \right)
                 \right)
\end{array}
\end{equation}
and 
\begin{equation} \label{eq:FS} \begin{array}{lll}
\D (fS^{\mu \nu}_a),_{\mu \nu} & \D  =  & \D  \frac{1}{v}
              \left\{ \frac{\partial^2 g}{\partial v^2} \left(vfS^{11}_a\right)
+2 \frac{\partial^2 g}{\partial v \partial v_{\varphi}} \left(vfS^{13}_a \right)
   \right. \\
& & \\
& & \left. \; \; \; \; \; 
  \D +\frac{\partial^2 g}{\partial v_{\varphi}^2} \left(vfS^{33  }_a \right)\right\} \\
& & \\
& & \left.
\D + \;  \frac{1}{v}\left\{ \frac{\partial}{\partial v} \left( vf \left[
       \frac{-(v^2-v_{\varphi}^2)}{v^3} \frac{\partial^2 g}{\partial 
                                                            v_{\varphi}^2}
        \right. \right.
        \right. \right.  \\
& & \\
& &        \left. \left. \left.  \; \; \; \; \; \; \; \; \; \; \; \; \; \; \;
                  \; \; \; \; \; \; \; \; \; \; \; \; \; \; \;
  \D      - \frac{2}{v^2} \frac{\partial g}{\partial v} \right] \right) \right\}
\label{eq:s22last}
\; .        \end{array}
\end{equation}
Most terms involving $S^{22}$ vanisch due to further derivations with respect
to the coordinate of symmetry, $\psi$, but one of them is retained in the
last term in Eq.~(\ref{eq:s22last}), therein written explicitly.
The diffusion coefficients in transformations between curvilinear
coordinate systems may now be identified using
\begin{equation} \label{eq:DV1}
      \frac{1}{\Gamma_a}
  <\Delta v^{\mu}>_a \: = \: T_a^{\mu} \: - \: \frac{1}{2}\Gamma^{\mu}_{\omega
                              \tau} S^{\omega \tau}
\end{equation}
\begin{equation} \label{eq:DVV1}
      \frac{1}{\Gamma_a}
  <\Delta v^{\mu} \Delta v^{\nu}> \: = \: S^{\mu \nu}_a \:\:,
\end{equation}
where the additional term  in Eq.~(\ref{eq:DV1}) with respect to the case
of Eq. (\ref{eq:DV2}) originates from the second (last) term of Eq. 
(\ref{eq:FS}).

It is convenient to treat the problem in energy - angular momentum - space
such that the diffusion coefficients just derived have to be transformed
to the new velocity variables $E=\frac{1}{2}v^2+\Phi(\rho,z)$ 
and $J_z=\rho v_{\varphi}$. This can be accomplished by using the following
simple formula:
\begin{equation} \label{eq:DEJ1}
  <\Delta E>  =      E,_{\mu}<\Delta v^{\mu}> + \frac{1}{2}E,_{\mu \nu}
                       <\Delta v^{\mu} \Delta v^{\nu}>  
\end{equation}
\begin{equation} 
  <\Delta J_z>  =    J,_{\mu}<\Delta v^{\mu}> + \frac{1}{2}J,_{\mu \nu}
                       <\Delta v^{\mu} \Delta v^{\nu}> 
\end{equation}
\begin{equation}
  <(\Delta E)^2>  =  E,_{\mu} E,_{\nu} <\Delta v^{\mu} \Delta v^{\nu}> 
\end{equation}
\begin{equation}
  <(\Delta J_z)^2>  =  J,_{\mu} J,_{\nu} <\Delta v^{\mu} \Delta v^{\nu}>
\end{equation}
\begin{equation}  \label{eq:DEJ2}
  <\Delta E \Delta J_z>  =  E,_{\mu}J,_{\nu}<\Delta v^{\mu} \Delta v^{\nu}>
\end{equation} 
If we now assume an isotropic background distribution $f(v_f)$, integrals
$h$ and $g$ are easily simplified ({\em cf.} Spitzer 1987, p.36). For this
purpose, we express the velocity $\vec{v}$ (inertial frame) in terms of 
velocities $(\vec{u}+\rho \Omega \vec{e_{\varphi}})$ in the corotating 
frame. $\Omega$ is the angular
velocity of the corotating frame. Thus, $f(u)$ is isotropic and the
derivatives of $h$ and $g$ with respect to $v$ and $v_{\varphi}$
 ($\frac{\partial h}{\partial v}, \frac{\partial h}{\partial v_{\varphi}}
,$ etc.) must be transformed to those with respect to $u$ only. For example,
$\frac{\partial h}{\partial v_{\varphi}} = -\frac{\rho \Omega}{u}
\frac{\partial h}{\partial u}$. Inserting eqs. (\ref{eq:TENB}) to 
(\ref{eq:TENE}) in eqs. (\ref{eq:DV1}), (\ref{eq:DVV1}) and employing these
results again in eqs. (\ref{eq:DEJ1}...\ref{eq:DEJ2}) under 
consideration of the transformation
to the corotating frame, we arrive at
\begin{equation} \begin{array}{lcl}
\D <\Delta E> &\D =& \D \left(u+\frac{J_z\Omega}{u}-\frac{\rho^2\Omega^2}{u}\right)
               \frac{\partial h}{\partial u} \\
& &\\
& &          \D + \: \frac{1}{2} \frac{\partial ^2 g}{\partial u^2} \\
& &\\
& &          \D  + \: \frac{1}{u} \frac{\partial g}{\partial u} \: , 
\end{array}
\end{equation}
\begin{equation} \begin{array}{lcl} 
\D <\Delta J_z> & \D = & \D \left(\frac{J_z}{u}-\frac{\rho^2\Omega}{u}\right)
               \frac{\partial h}{\partial u} \: ,
\end{array}
\end{equation}
\begin{equation} \begin{array}{lcl}
\D <(\Delta E)^2> &\D = & \D \left(u^2+2J_z\Omega-2\rho^2 \Omega^2 
                     +\frac{J_z^2\Omega^2}{u^2} \right. \\
& & \\
& &              \left. \;\;\; \D   -\frac{2J_z\rho^2\Omega^3}{u^2}
                      +\frac{\rho^4\Omega^4}{u^2}\right)
                \D   \frac{\partial ^2 g}{\partial u^2} \\
& & \\
& &         \D + \: \left( \frac{\rho^2 \Omega^2}{u}
                     -\frac{J_z^2\Omega^2}{u^3} \right. \\
& & \\
& &           \left. \;\;\;\;\;\; \D   +\frac{2J_z\rho^2\Omega^3}{u^3} 
                     -\frac{\rho^4\Omega^4}{u^3}\right)
               \D    \frac{\partial g}{\partial u}  \: ,
\end{array}
\end{equation}
\begin{equation} \begin{array}{lcl}
\D <(\Delta J_z)^2> & \D = & \D \left( \frac{J_z^2}{u^2}+\frac{\rho^4\Omega^2}{u^2}
                       -\frac{2J_z\rho^2\Omega}{u^2}\right)
                     \frac{\partial ^2 g}{\partial u^2} \\
& &\\
& &     \D    + \: \left( \frac{-J_z^2}{u^3}-\frac{\rho^4\Omega^2}{u^3}\right. \\
& & \\
& &              \left.  \;\;\;\;\;\;
         \D          +\frac{\rho^2}{u}+\frac{2J_z\rho^2\Omega}{u^3}\right)
                   \D  \frac{\partial g}{\partial u} \: ,
\end{array}
\end{equation}
\begin{equation} \begin{array}{lcl}
\D <\Delta E \Delta J_z> & \D = & \D \left( J_z-\rho^2\Omega
                      +\frac{J_z^2\Omega}{u^{2}} \right. \\
& & \\
& &                   \left. \;\;\; \D -\frac{2J_z\rho^2\Omega^2}{u^2}
                       +\frac{\rho^4\Omega^3}{u^2}\right)
                     \D \frac{\partial ^2 g}{\partial u^2} \\
& & \\
& &        \D  + \: \left( \frac{\rho^2\Omega}{u} + \frac{2J_z\rho^2\Omega^2}{u^3}
                 \right. \\
& & \\
& &   \left. \;\;\;\;\;\;
      \D     -\frac{J_z^2\Omega}{u^3}-\frac{\rho^4\Omega^3}{u^3}\right)
                  \D   \frac{\partial g}{\partial u}\: , 
\end{array} \end{equation}
The Fokker-Planck equation is usually recast in flux conservation form:
\begin{equation}\label{eq:FX1} 
      \frac{Df}{Dt} = -\frac{\partial F_E}{\partial E}
                      -\frac{\partial F_{J_z}}{\partial J_z},
\end{equation}
where we still have not applied an orbit average. The term on the left hand
side represents the Vlasov-term of the full collisional Boltzmann equation. 
The fluxes $F$ are given by
\begin{equation} \label{eq:FX2}
    F_E = -D_{EE}\frac{\partial f}{\partial E}
          -D_{EJ_z}\frac{\partial f}{\partial J_z}
          -D_{E} f
\end{equation}
\begin{equation} \label{eq:FX3}
    F_{J_z} = -D_{J_zJ_z}\frac{\partial f}{\partial J_z}
              -D_{J_zE}\frac{\partial f}{\partial E}
              -D_{J_z} f .
\end{equation} 
Comparing eqs. (\ref{eq:FX1} - \ref{eq:FX3}) with the original form (e.g.
Spitzer 1987) the flux coefficients D may be identified to
\begin{equation} \begin{array}{lcl}
D_{EE} & = & \frac{1}{2}<(\Delta E)^2> \\
& & \\
D_{EJ_z} & = & \frac{1}{2}<\Delta E \Delta J_z> \\
& & \\
D_{J_zJ_z} & = & \frac{1}{2}<(\Delta J_z)^2> \\
& & \\
D_{JE_z} & = & \frac{1}{2}<\Delta E \Delta J_z> \\
& & \\
D_{E} & = & -<\Delta E> + \frac{1}{2}\frac{\partial}{\partial E} 
            <(\Delta E)^2> \\
& & \\
& &         \;\;\;\;\;\;\;\;\;\;\;\;\;\;\;\;\;\; 
             +\frac{1}{2}\frac{\partial}{\partial J_z}
            <\Delta E \Delta J_z> \\
& & \\
D_{J_z} & = & -<\Delta J_z> +\frac{1}{2}\frac{\partial}{\partial J_z}
             <(\Delta J_z)^2> \\
& & \\
& &         \;\;\;\;\;\;\;\;\;\;\;\;\;\;\;\;\;\; 
              + \frac{1}{2}\frac{\partial}{\partial E}
             <\Delta E \Delta J_z>
\end{array} 
\end{equation}
Inserting the expressions for the diffusion coefficients given above,
one obtains finally the desired flux coefficients for the axially symmetric
application with rotation. We find
\begin{equation} \begin{array}{lcl}
\D D_{E}  & \D = & \D 4\pi
     \left( \frac{J_{z}\Omega}{u}-\frac{\rho^{2}\Omega^{2}}{u}
                    +u\right) F_{2}(u),
\end{array} \end{equation}
\begin{equation} \begin{array}{lcl}
\D D_{J_{z}} & \D = & \D 4\pi
     \left( \frac{J_{z}}{u} - \frac{\rho^{2}\Omega}{u}
                \right) F_{2}(u),
\end{array} \end{equation}
\begin{equation} \begin{array}{lcl}
\D D_{EE} & \D = & \D \frac{4\pi}{3}\left( 2uJ_{z}\Omega - u\rho^{2}\Omega^{2}
                                            + u^{3} \right) E_{1}(u) \\
       &   &  \\
       &   & 
             \D +2\pi \left( u\rho^{2}\Omega^{2} - \frac{J_{z}^{2}
               \Omega^{2}}{u} \right. \\
       &   &  \\
       &   &  \left. \;\;\;\;\;\;\;\;\;\;\;\;\;\;\;\;\;\;
          \D     + \frac{2J_{z}\rho^{2}\Omega^{3}}{u}
               -\frac{\rho^{4}\Omega^{4}}{u}\right) F_{2}(u) \\
       &   &  \\
       &   &  \D  + 2\pi \left( \frac{2u^3}{3} + \frac{4}{3}uJ_{z\rho}
                  - \frac{5}{3}u\rho^{2}\Omega^{2} \right. \\
       &   &  \\
       &   &  \left. \;\;\;\;\;\;\;\;\;\;\;\;\;\;\;\;\;\; 
              \D    + \frac{J_{z}^2\Omega^{2}}{u}
                  + \frac{\rho^{4}\Omega^{4}}{u}\right) F_{4}(u) \; ,
\end{array} \end{equation}
\begin{equation} \begin{array}{lcl}
\D D_{J_{z}J_{z}} & \D = &  \D \frac{4\pi}{3}u\rho^2 E_{1}(u)  \\
& & \\
& &    \D      + 2\pi\left( u\rho^2 - \frac{J_z^2}{u} 
               - \frac{\rho^4\Omega^2}{u} \right. \\
& & \\
& &         \left. \;\;\;\;\;\;\;\;\;\;\;\;\;\;\;\;\;\;\;\;\;
        \D     + \frac{2J_z\rho^2\Omega}{u} \right) 
                F_{2}(u) \\
& & \\
& &      \D     2\pi \left( \frac{J_{z}^{2}}{u} 
             + \frac{\rho^4\Omega^2}{u} \right. \\
& & \\
& &          \left. \;\;\;\;\;\;\;\;\;\;\;\;\;\;\;\;\;\;\;\;\;
          \D     - \frac{2J_{z}\rho^{2}\Omega}{u}
             - \frac{1}{3}u\rho^2\right)
                F_{4}(u) \; .
\end{array} 
\end{equation}
\begin{equation} \begin{array}{lcl}
\D D_{EJ_{z}} & \D = & \D \frac{4\pi}{3}uJ_{z} E_{1}(u) \\
& & \\
& & \D + 2\pi \left( u\rho^{2}\Omega + \frac{2J_{z}\rho^{2}\Omega^{2}}
                      {u} \right. \\
& & \\
& &               \left. \;\;\;\;\;\;\;\;\;\;\;\;\;\;\;\;\;\;\;\
          \D        \;\; - \frac{J_{z}^{2}\Omega}
                      {u} - \frac{\rho^4\Omega^3}{u} \right) F_{2}(u) \\
& & \\
& &  \D
+ 2\pi \left( \frac{2}{3}uJ_{z} - u\rho^{2}\Omega
                      - \frac{2J_{z}\rho^{2}\Omega^{2}}{u} \right. \\
& & \\
& &                \left. \;\;\;\;\;\;\;\;\;\;\;\;\;\;\;\;\;\;\;\; +
            \D      \frac{\rho^4\Omega^3}{u} + \frac{J_{z}^{2}\Omega}
                       {u} \right) F_{4}(u) \; .
\end{array}
\end{equation}
The functions $E_{i}$ and $F_{i}$ are constituent parts of the Rosenbluth
potentials and their derivatives. 
\begin{equation} 
    F_{i} = \frac{1}{u^{i}} \int_0^u u'^i f(u')du'
\end{equation}
\begin{equation}
    E_{i} = \frac{1}{u^{i}} \int_u^\infty u'^i f(u')du'
\end{equation}
Given an expression for the background distribution function (section 
\ref{sec:mod}, rotating King-model used throughout this paper) 
these functions are evaluated for each position in the 
meridional plane,
when the local density, mean particle velocity and mean particle angular
velocity are specified. 

The next step is to orbit average these flux coefficients and again to 
transform them to ($X,Y$)-coordinates used throughout our numerical
simulations.
\label{lastpage}

\end{document}

\bsp

\label{lastpage}

\end{document}